\documentclass[twocolumn,showpacs,amsmath,pre]{revtex4}

\usepackage{graphicx}
\usepackage{amsmath}
\usepackage{amssymb}
\usepackage{natbib}
\usepackage{lscape}
\usepackage{rotating}

\begin{document}

\title{Reconstruction of Ordinary Differential Equations From Time Series Data}

\author{Manuel Mai$^{1}$}
\author{Mark D. Shattuck$^{2}$}
\author{Corey S. O'Hern$^{3,1,4,5}$}

\affiliation{$^{1}$ Department of Physics, Yale University, New Haven,
  Connecticut 06520, USA}

\affiliation{$^{2}$ Benjamin Levich Institute and Physics Department, The City College of New York, New York, New York 10031, USA}

\affiliation{$^{3}$ Department of Mechanical Engineering \& Materials
  Science, Yale University, New Haven, Connecticut 06520, USA}

\affiliation{$^{4}$ Department of Applied Physics, Yale University, New Haven, Connecticut 06520, USA}

\affiliation{$^{5}$ Graduate Program in Computational Biology and Bioinformatics, Yale University, New Haven, Connecticut 06520, USA}

\begin{abstract}
We develop a numerical method to reconstruct systems of ordinary
differential equations (ODEs) from time series data without {\it a
  priori} knowledge of the underlying ODEs using sparse basis learning
and sparse function reconstruction.  We show that employing sparse
representations provides more accurate ODE reconstruction compared to
least-squares reconstruction techniques for a given amount of time
series data. We test and validate the ODE reconstruction method on
known 1D, 2D, and 3D systems of ODEs. The 1D system possesses two
stable fixed points; the 2D system possesses an oscillatory fixed
point with closed orbits; and the 3D system displays chaotic dynamics
on a strange attractor.  We determine the amount of data required to
achieve an error in the reconstructed functions to less than $0.1\%$.
For the reconstructed 1D and 2D systems, we are able to match the
trajectories from the original ODEs even at long times.  For the 3D
system with chaotic dynamics, as expected, the trajectories from the
original and reconstructed systems do not match at long times, but the
reconstructed and original models possess similar Lyapunov
exponents. Now that we have validated this ODE reconstruction method
on known models, it can be employed in future studies to identify new
systems of ODEs using time series data from deterministic systems for
which there is no currently known ODE model.
  
\end{abstract}


\maketitle

\section{Introduction}
\label{intro}

We will present a methodology to create ordinary differential
equations (ODEs) that reproduce measured time series data from
physical systems. In the past, physicists have constructed ODEs by
writing down the simplest mathematical expressions that are consistent
with the symmetries and fixed points of the system. For example,
E. N. Lorenz developed an ODE model for atmospheric
convection~\cite{lorenz1963} by approximating solutions to the
Navier-Stokes equations for Rayleigh-B\'{e}nard convection. Beyond its
specific derivation, the Lorenz system of ODEs is employed to model a
wide range of systems that display nonlinear and chaotic dynamics,
including lasers~\cite{weiss1986}, electrical
circuits~\cite{cuomo1993}, and MEMS~\cite{aref2002} devices.

ODE models are also used extensively in computational biology. For
example, in systems biology, genetic circuits are modeled as networks
of electronic circuit elements~\cite{Csete2002, villaverde2014}.  In
addition, systems of ODEs are often employed to investigate viral
dynamics ({\it e.g.} HIV \cite{perelson1999, callaway2002,nelson2002},
hepatitis \cite{dahari2009,gourley2008}, and influenza
\cite{hancioglu2007,perelson2002}) and the immune system response to
infection~\cite{day1,day2}. Population dynamics and epidemics have
also been successfully modeled using systems of ODEs
\cite{arditi1989}.  In most of these cases, an {\it ad hoc} ODE model
with several parameters is posited~\cite{baake1992}, and solutions of
the model are compared to experimental data to identify the relevant
range of parameter values.

A recent study has developed a more systematic computational approach
to identify the ``best'' ODE model to recapitulate time series data.  The
approach iteratively generates random mathematical expressions for a 
candidate ODE model. At each iteration, the ODE model is solved and
the solution is compared to the time series data to identify the
parameters in the candidate model.  The selected parameters minimize
the distance between the input trajectories and the solutions of the
candidate model.  Candidate models with small errors are then co-evolved
using a genetic algorithm to improve the fit to the input time series
data~\cite{simeone2006,bongard2007,schmidt2008,schmidt2009}. The
advantage of this method is that it yields an approximate analytical expression
for the ODE model for the dynamical system. The disadvantages of this approach
include the computational expense of repeatedly solving
ODE models and the difficulty in finding optimal solutions
for multi-dimensional nonlinear regression.

Here, we develop a method to build numerical expressions of a system
of ODEs that will recapitulate time series data of a dynamical
system. This method has the advantage of not needing any input except
the time series data, although {\it a priori} information about the
fixed point structure and basins of attraction of the dynamical system
would improve reconstruction.  Our method includes several steps.  We
first identify a basis to sparsely represent the time series data
using sparse dictionary learning
\cite{olshausen1997,aharon2006,aharon2008}. We then find the
sparsest expansion in the learned basis that is consistent with the
measured data. This step can be formulated as solving an
underdetermined system of linear equations. We will solve the
underdetermined systems using ${\rm L}_1$-norm regularized regression, which
finds the solution to the system with the fewest nonzero expansion
coefficients in the learned basis.  We test our ODE reconstruction
method on time series data generated from known ODE models in one-,
two-, and three-dimensional systems, including both non-chaotic 
and chaotic dynamics.  We quantify the accuracy of the
reconstruction for each system of ODEs as a function of the amount of
data used by the method. Further, we solve the reconstructed
system of ODEs and compare the solution to the original time series
data.  The method developed and validated here can now be applied to
large data sets for physical and biological systems for which there is
no known system of ODEs.

Identifying sparse representations of data ({\it i.e.} sparse coding)
is well studied. For example, sparse coding has been widely used for
data compression, yielding the JPEG, MPEG, and MP3 data formats.
Sparse coding relies on the observation that for most signals a basis
can be identified for which only a few of the expansion coefficients
are nonzero~\cite{smith2006,raina2007,mairal2009supervised}. Sparse
representations can provide accurate signal recovery, while at the
same time, reduce the amount of information required to define the
signal. For example, keeping only the ten largest coefficients out of
64 possible coefficients in an $8\times8$ two-dimensional discrete
cosine basis (JPEG), leads to a size reduction of approximately a
factor of $6$.

Recent studies have shown that in many cases perfect recovery of a
signal is possible from only a small number of measurements of the
signal~\cite{candes2006,donoho2006,candes2006cs,donoho2006cs,candes2006rb}. This work provided a new lower bound for
the amount of data required for perfect reconstruction of a signal;
for example, in many cases, one can take measurements at frequencies
much below the Nyquist sampling rate and still achieve perfect signal
recovery.  The related field of compressed sensing emphasizes sampling
the signal in compressed form to achieve perfect signal reconstruction
\cite{candes2006cs, donoho2006cs,candes2006rb,candes2006stable,baraniuk2007,candes2008,donoho2010}.
Compressed sensing has a wide range of applications from speed up
of magnetic resonance image
reconstruction~\cite{lustig2007,lustig2008,haldar2011}
to more efficient and higher resolution cameras~\cite{duarte2008,yang2010}.

Our ODE reconstruction method relies on the assumption that the
functions that comprise the ``right-hand sides'' of
the systems of ODEs can be sparsely represented in some basis. A function
$f({\vec x})$ can be sparsely represented by a set of basis functions
$\{\phi_i\},\ i=1,\dots,n$ if $f({\vec x}) = \sum_{i=1}^n c_i
\phi_i({\vec x})$ with only a small number $s \ll n$ of nonzero
coefficients $c_i$. This assumption is not as restrictive
as it may seem at first. For example, suppose we sample a
two-dimensional function on a discrete $128\times128$ grid. Since
there are $128^2=16384$ independent grid points, a complete basis
would require at least $n=16384$ basis functions. For most
applications, we expect that a much smaller set of basis functions
would lead to accurate recovery of the function. In fact, the sparsest
representation of the function is the basis that contains the function
itself, where only one of the coefficients $c_i$ is nonzero. Identifying
sparse representations of the system of ODEs is
also consistent with the physics paradigm of finding the simplest
model to explain a dynamical system.
 
The remainder of this manuscript is organized as follows. In the
methods section (Sec.~\ref{methods}), we provide a formal definition
of sets of ODEs and details about obtaining the right-hand side
functions of ODEs from numerically differentiating time series data.
We then introduce the concept of $L_1$ regularized regression and
apply it to the reconstruction of a sparse undersampled signal. We
introduce the concept of sparse basis learning to identify a basis in
which the ODE can be represented sparsely. At the end of the methods
section, we define the error metric that we will use to quantify the
accuracy of the ODE reconstruction. In the results section
(Sec.~\ref{ofd_Results}), we perform ODE reconstruction on models in
one-, two-, and three-dimensional systems. For each system, we measure
the reconstruction accuracy as a function of the amount of data that
is used for the reconstruction, showing examples of both accurate and
inaccurate reconstructions. We end the manuscript in
Sec.~\ref{discussion} with a summary and future applications
of our method for ODE reconstruction.

\section{Methods}
\label{methods}

In this section, we first introduce the mathematical expressions that
define sets of ordinary differential equations (ODEs). We then
describe how we obtain the system of ODEs from time
series data.  In Secs.~\ref{sparse_coding} and~\ref{basis_learning},
we present the sparse reconstruction and sparse basis learning methods
that we employ to build sparse representations of the ODE model. We
also compare the accuracy of sparse versus non-sparse methods for
signal reconstruction. In Sec.~\ref{models}, we introduce the specific one-,
two-, and three-dimensional ODE models that we use to validate our
ODE reconstruction methods.

\subsection{Systems of ordinary differential equations}
\label{expression}

A general system of $N$ nonlinear ordinary differential equations
is given by
\begin{equation}
\label{eq:odesystem}
\begin{aligned}
\frac{dx_1}{dt}&=f_1(x_1,x_2,\dots,x_N)\\
\frac{dx_2}{dt}&=f_2(x_1,x_2,\dots,x_N)\\
&\vdots\\
\frac{dx_N}{dt}&=f_n(x_1,x_2,\dots,x_N),
\end{aligned}
\end{equation}
where ${f_i}$ with $i=1,\ldots,N$ are arbitrary nonlinear functions of
all $N$ variables $x_i$ and $d/dt$ denotes a time derivative.
Although the functions $f_i(x_1,\ldots,x_N)$ are defined for all
values of $x_i$ within a given domain, the time derivatives of the
solution ${\vec x}(t)=(x_1(t),x_2(t),\dots,x_N(t))^T$ only sample the
functions along particular trajectories. The functions $f_i$ can be
obtained by taking numerical derivatives of the solutions with respect
to time:
\begin{equation}
\label{eq:fsample}
f_i (t_0)\approx \frac{x_i(t_0+\Delta t) -x_i(t_0) }{\Delta t}. 
\end{equation}
We will reconstruct the functions from a set of $m$ measurements,
$y_k=f_i(\vec{x}^k)$ at positions $\{\vec{x}^k\}$ with
$k=1,\ldots,m$. To do this, we will express the functions $f_i$
as linear superpositions of arbitrary, non-linear basis functions
$\phi_j(\vec{x})$:
\begin{equation}
\label{eq:f_expansion}
f_i(\vec x) = \sum_{j=1}^{n} c_j\,\phi_j(\vec x),
\end{equation} 
where $c_j$ are the expansion coefficients and $n$ is the number of
basis functions used in the expansion. The measurements $y_k =
f_i(\vec x^k)$ impose the following constraints on the expansion coefficients:
\begin{equation}
\label{eq:c_conditions}
	f_i(\vec x^k) = y_k=\sum_{j=1}^n c_j\,\phi_j(\vec x^k)
\end{equation}
for each $k=1,\dots,m$. The constraints in Eq.~\ref{eq:c_conditions}
can also be expressed as a matrix equation: $\sum_{j=1}^n
\Phi_{ij} c_{j} = y_{i}$ for $i=1,\ldots,m$ or 
\begin{equation}
\label{eq:matrixeq}
\begin{pmatrix}
\phi_1({\vec x}^1) & \hdots & \phi_n({\vec x}^1)\\
\phi_1({\vec x}^2) & \hdots & \phi_n({\vec x}^2)\\
&\vdots&\\
\phi_1({\vec x}^m) & \hdots & \phi_n({\vec x}^m)
\end{pmatrix}
\begin{pmatrix}
c_1\\c_2\\ \vdots\\c_n
\end{pmatrix}
=
\begin{pmatrix}
y_1 \\ y_2\\ \vdots \\ y_m
\end{pmatrix},	
\end{equation}
where $\Phi_{ij} = \phi_j({\vec x}^i)$.  In most cases of ODE reconstruction,
the number of rows of $\Phi$, {\it i.e.}  the number of measurements
$m$, is smaller than the number of columns of $\Phi$, which is equal
to the number of basis functions $n$ used to represent the signals
$f_i$. Thus, in general, the system of equations
(Eq.~\ref{eq:matrixeq}) is underdetermined with $n>m$.

Sparse coding is ideally suited for solving underdetermined systems
because it seeks to identify the minimum number of basis functions to
represent the signals $f_i$. If we identify a basis that can represent
a given set of signals $f_i$ sparsely, an $L_1$ regularized
minimization scheme will be able to find the sparsest representation
of the signals~\cite{donohol1}.

\subsection{Sparse Coding}
\label{sparse_coding}

In general, the least squares ($L_2$) solution of
Eq.~\ref{eq:matrixeq} possesses many nonzero coefficients
$c_i$, whereas the minimal $L_1$ solution of Eq.~\ref{eq:matrixeq} is
sparse and possesses only a few non-zero coefficients. In the case of
underdetermined systems with many available basis functions, it has
been shown that a sparse solution obtained via $L_1$ regularization
more accurately represents the solution compared to those that 
are superpositions of many basis functions~\cite{donohol1}.

The solution to Eq.~\ref{eq:matrixeq} can be obtained by minimizing
the squared differences between the measurements of the signal ${\vec
  y}$ and the reconstructed signal $\Phi {\vec c}$ subject to the
constraint that the solution ${\hat c}$ is sparse~\cite{scikit-learn}:
\begin{equation}
\label{eq:L1min}
\hat c = \min_{{\vec c}} \frac{1}{2} \| {\vec y} - \Phi {\vec c} \|_2^2 + \lambda \| {\vec c}\|_1,
\end{equation} 
where $\|.\|_p$ denotes the $L_p$ vector norm and
$\lambda$ is a Langrange multiplier that penalizes a large $L_1$ norm
of ${\hat c}$. The $L_p$ norm of an
$n$-dimensional vector $\vec{x}$ is defined as
\begin{equation}
\| \vec{x}\|_p = \left[\sum_{i=1}^N|x_i|^p\right]^{1/p}
\end{equation}
for $p>1$, where $N$ is the number of components of the vector ${\vec x}$.

We now demonstrate how the sparse signal reconstruction method
compares to a standard least squares fit. We first construct a sparse
signal (with sparsity $s$) in a given basis. We then sample the signal
randomly and attempt to recover the signal using the regularized
$L_1$ and least-squares reconstruction methods. For this example, we
choose the discrete cosine basis.  For a signal size of $100$ values, we have 
a complete and orthonormal basis of $100$ functions $\phi_n(i)$
($n=0,\ldots,99$) each with $100$ values ($i=0,\ldots,99$):
\begin{equation}
\label{eq:dct}
 	\phi_n(i) = F(n)\,{\rm cos}\left[\frac{\pi}{100}\left(i+\frac12\right)n\right],
 \end{equation} 
where $F(n)$ is a normalization factor
\begin{equation}
\label{eq:dct_norm}
F(n)=
\begin{cases}
\frac{1}{\sqrt{100}} & {\rm for}\quad n = 0\\
\sqrt{\frac{2}{100}} & {\rm for}\quad n = 1,\dots,99.
\end{cases}	
\end{equation}
Note that an orthonormal basis is not
a prerequisite for the sparse reconstruction method. 

Similar to Eq.~\ref{eq:f_expansion}, we can express the signal as
a superposition of basis functions,
\begin{equation}
\label{eq:signal}
g(i)=\sum_{j=0}^{99}c_j\phi_j(i)\,.
\end{equation}
The signal $\vec{g}$ of sparsity $s$ is generated by
randomly selecting $s$ of the coefficients $c_j$ and assigning them a
random amplitude in the range $[-1,1]$.  We then evaluate $g(i)$ at
$m$ randomly chosen positions $i$ and attempt to recover $g(i)$ from
the measurements. If $m<100$, recovering the original signal involves
solving an underdetermined system of linear equations.  

Recovering the full signal ${\vec g}$ from a given number of
measurements proceeds as follows. After carrying out the $m$
measurements, we can rewrite Eq.~\ref{eq:c_conditions} as
\begin{equation}
	\vec{y} = P \vec{g},
\end{equation}
where $\vec{y}$ is the vector of the measurements of $\vec{g}$ and $P$
is the projection matrix with $m\times n$ entries that are either $0$
or $1$.  Each row has one nonzero element that corresponds to the
position of the measurement. For each random selection of measurements of $\vec{g}$,
we solve the reduced equation
\begin{equation}
\label{eq:underdeteqs}
	\vec{ y} = \Theta \vec{ c}\,,
\end{equation}
where $\Theta = P \Phi$. After solving Eq.~\ref{eq:underdeteqs} for
$\hat {c}$, we obtain a reconstruction of the original signal
\begin{equation}
	\label{eq:frec}
	\vec{g}_{rec} = \Phi\,\hat {\mathbf c}\,.
\end{equation}

\begin{figure}
\includegraphics[width=3.5in]{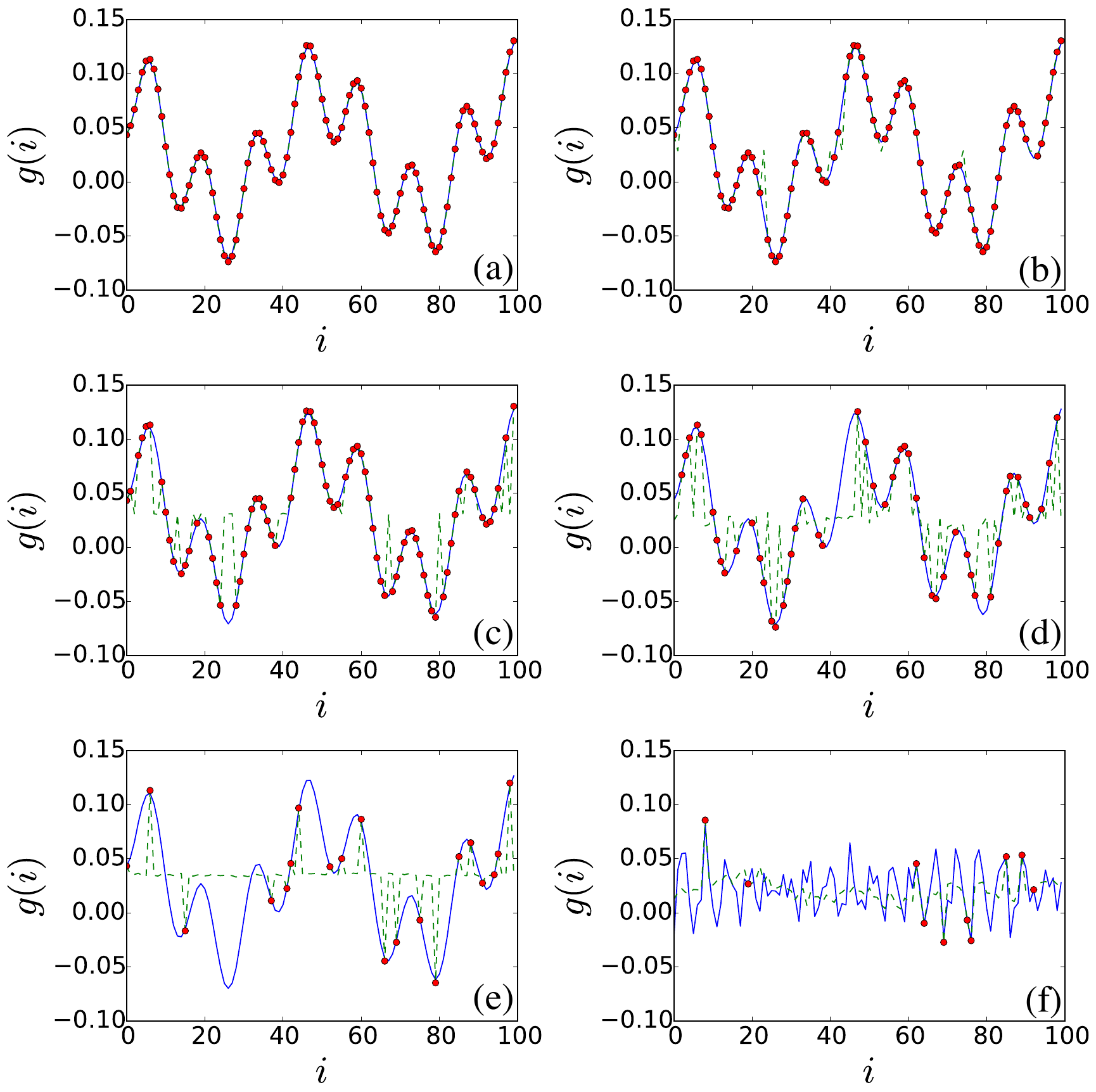}
\caption{Comparison of the least squares $L_2$ (green dashed lines)
and $L_1$ (blue solid lines) regularized regression to recover a
randomly generated, discrete signal (red dots in panel (a)) obtained
by randomly sampling points $i$ from the function $g(i)=
\sum_{n=0}^{99} c_n \phi_n(i)$, where $\phi_n(i)$ is given by
Eq.~\ref{eq:dct} and $i = 0,\ldots,99$ is chosen so that all $n$
frequencies can be resolved. The function was constructed to be 3
sparse ($s=3$) with only $c_{0}$, $c_{4}$, and $c_{15} \ne 0$.  The
six panels (a)-(f) show the reconstructions for $M = 1$, $0.9$,
$0.8$, $0.5$, $0.2$, and $0.1$ of the input signal, respectively,
included in the measurements. The red dots in panels (a)-(f)
indicate the fraction of the signal that was measured and used for
the reconstruction.}
\label{fi:intro_L1_L2_example}
\end{figure}

Fig.~\ref{fi:intro_L1_L2_example} shows examples of $L_1$ and $L_2$
reconstruction methods of a signal as a function of the fraction of 
the signal ($M=0.2$ to $1$) included in the reconstruction method.
Even when only a small fraction of the signal is included (down to
$M=0.2$), the $L_1$ reconstruction method achieves nearly perfect
signal recovery. In contrast, the least-squares method only
achieves adequate recovery of the signal for $M>0.9$. Moreover, when only a
small fraction of the signal is included, the $L_2$ method is dominated by the
mean of the measured points and oscillates rapidly about the mean to
match each measurement.  

\begin{figure}
\includegraphics[width=2.5in]{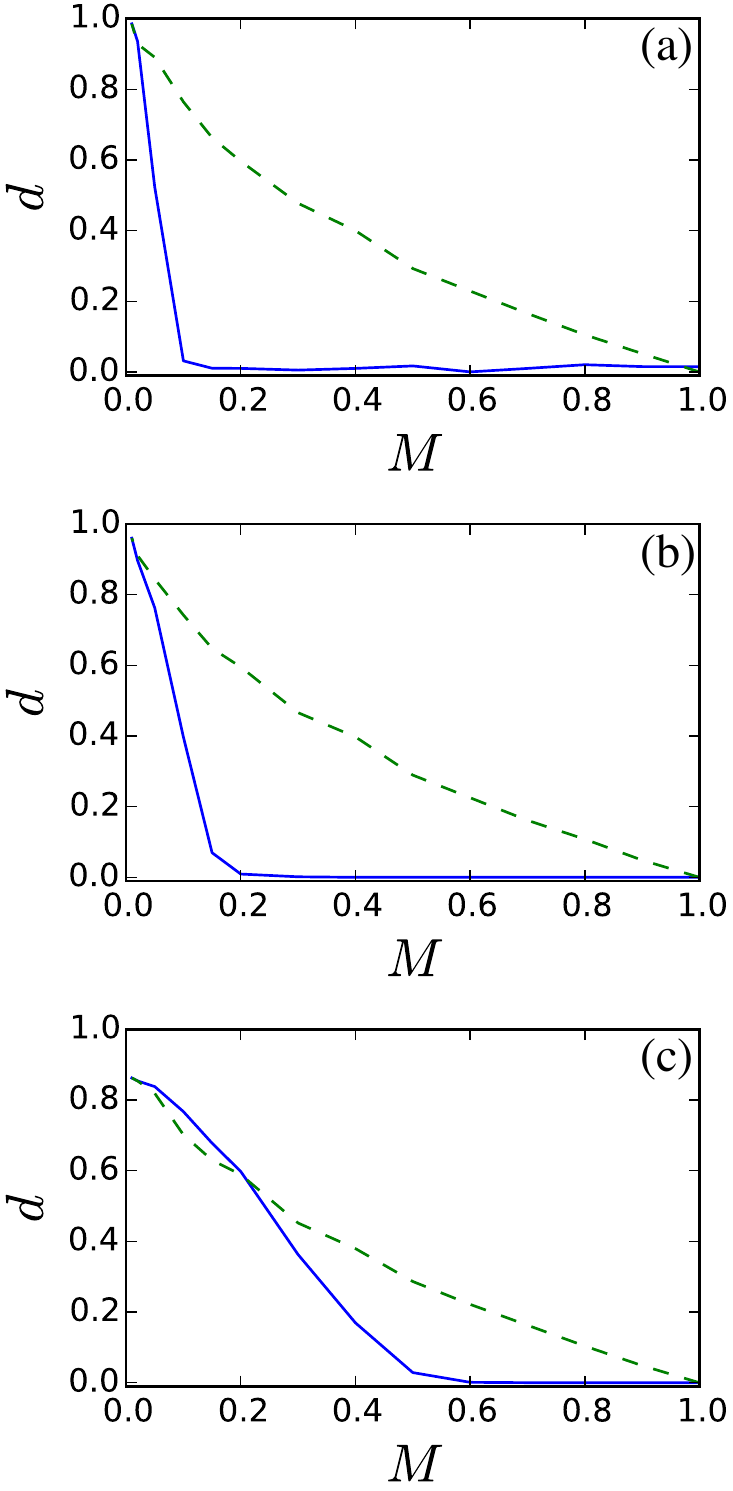}

\caption{Comparison of the least-squares $L_2$
(green dashed lines) and $L_1$ (solid blue lines) reconstruction errors $d$ 
(Eq.~\ref{eq:distance}) as a function of the fraction $M$ of the input 
signal $g(i)$ in Fig.~\ref{fi:intro_L1_L2_example} for three values of the 
sparsity (a) $s=1$, (b) $3$, and (c) $20$. 
}
\label{fi:intro_L1_L2_s}
\end{figure}

In Fig.~\ref{fi:intro_L1_L2_s}, we measured the recovery error $d$
between the original (${\vec g}$) and recovered (${\vec g}_{rec}$) signals
as a function of the fraction $M$ of the signal included and for several
sparsities $s$. We define the recovery error as 
\begin{equation}
d({\vec g},{\vec g}_{rec}) = 1-\frac{ {\vec g}\cdot {\vec g}_{rec} } {\|{\vec g}\|_2\,\|{\vec g}_{rec}\|_2}\,,
\label{eq:distance}
\end{equation}
where ${\vec g} \cdot {\vec g}_{rec}$ denotes the inner product
between the two vectors ${\vec g}$ and ${\vec g}_{rec}$.  This distance
function satisfies $0 \le d \le 2$, where $d\ge 1$ signifies a large
difference between ${\vec g}$ and ${\vec g}_{rec}$ and $d=0$ indicates
${\vec g} = {\vec g}_{rec}$.

For sparsity values $s=1$ and $3$, the $L_1$ reconstruction gives
small errors ($d \sim 0$) for $M\ge 0.2$
(Fig.~\ref{fi:intro_L1_L2_s}).  In contrast, the error for the $L_2$
reconstruction method is nonzero for all $M < 1$ for all $s$.  For a
non-sparse signal ($s=20$), the $L_1$ and $L_2$ reconstruction methods
give similar errors for $M \lesssim 0.2$.  In this case, the
measurements truly undersample the signal, and thus providing less
than $20$ measurements is not enough to constrain the $20$ nonzero
coefficients $c_j$. However, when $M \gtrsim 0.2$, $d$ from the $L_1$
reconstruction method is less than that from the $L_2$ method and is
nearly zero for $M \gtrsim 0.5$.

\subsection{Sparse Basis Learning}
\label{basis_learning}

The $L_1$ reconstruction method described in the previous section
works well if 1) the signal has a sparse representation in some basis
and 2) the basis $\Phi$ (or a subset of it) contains functions similar
to the basis in which the signal is sparse. How do we proceed with
signal reconstruction if we do not know a basis in which the signal is
sparse?  One method is to use one of the common basis sets, such as
wavelets, sines, cosines, or
polynomials~\cite{crutchfield1987,judd1995,judd1998}.  Another method
is to employ {\it sparse basis learning} that identifies a basis in
which a signal can be expressed sparsely. This approach is compelling
because it does not require significant prior knowledge about the
signal and it allows the basis to be learned even from noisy or
incomplete data.

Sparse basis learning seeks to find a basis $\Phi$ that can represent an
input of several signals sparsely. We identify $\Phi$ by decomposing
the signal matrix ${Y=(\vec{y}_1, \vec{y}_2, \dots, \vec{y}_m})$,
where $m$ is the number of signals, into the basis matrix $\Phi$ and
coefficient matrix $C$,
\begin{equation}
	Y = \Phi C\,.
\end{equation}
Columns $\vec{c}_i$ of $C$ are the sparse coefficient vectors that
represent the signals $\vec{y}_i$ in the basis $\Phi$. Both $C$ and
$\Phi$ are unknown and can be determined by minimizing the squared
differences between the signals and their representations in the basis
$\Phi$ subject to the constraint that the coefficient matrix is
sparse~\cite{mairal2009,scikit-learn}:
\begin{equation}
\label{eq:basis_learning_ch}
	\hat C,\hat \Phi = \min_{C,\Phi} \sum_{i=1}^{m}\left({\frac{1}{2} \| {\vec y}_i - \Phi {\vec c}_i \|_2^2 + \lambda \|{\vec c}_i\|_1}\right)\,,
\end{equation}
where $\lambda$ is a Langrange multiplier that determines 
the sparsity of the coefficient matrix $C$. 

\begin{figure}
\includegraphics[width=3in]{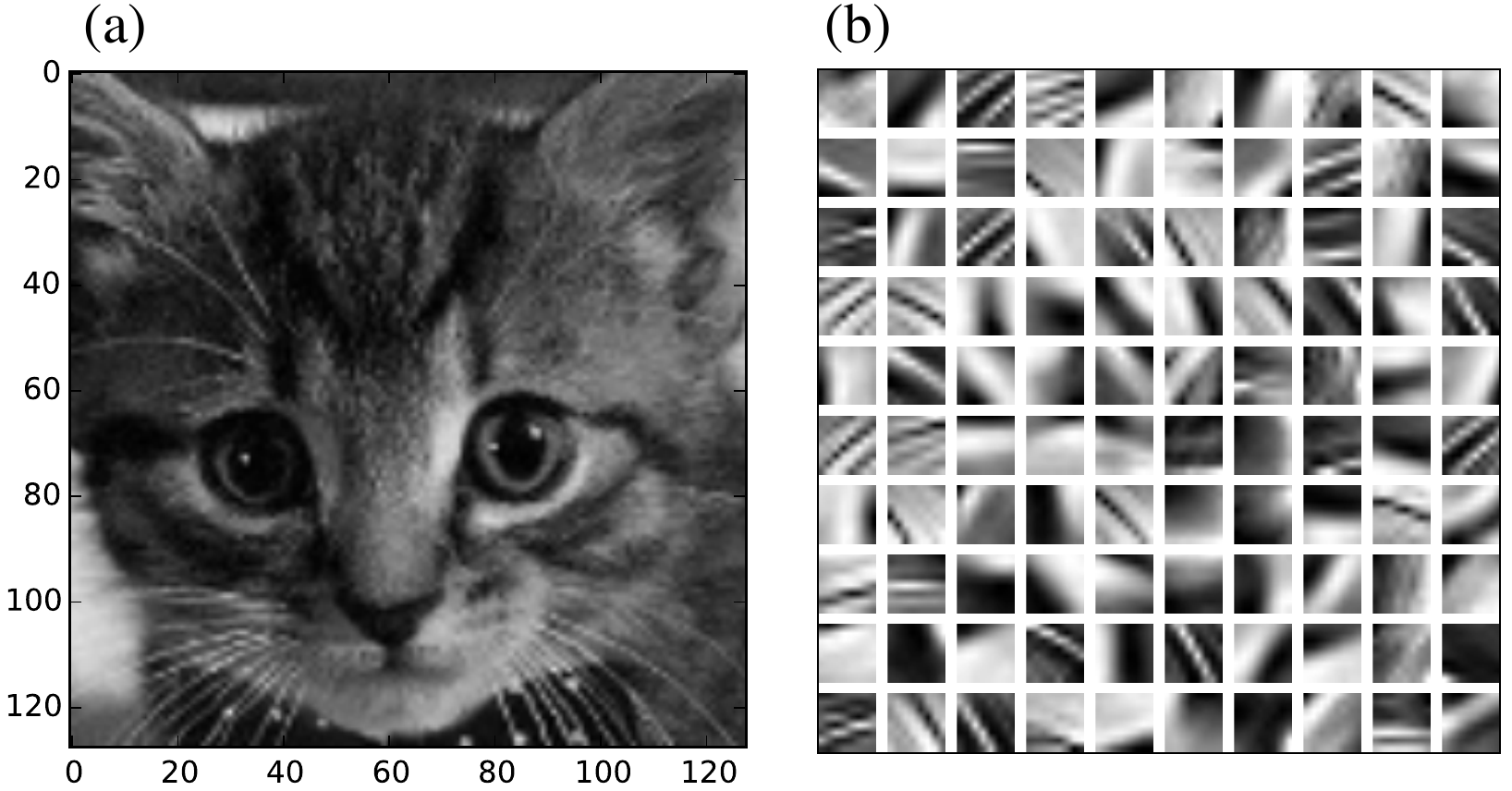}
\caption{(a) $128\times 128$ pixel image of a cat. (b) Using the
sparse learning method, we obtained $100$ $8\times8$ pixel basis
functions for the image in (a).}
\label{fi:cat_basis}
\end{figure}

To illustrate the basis learning method, we show the results of sparse
basis learning on the complex, two-dimensional image of a cat shown in
Fig.~\ref{fi:cat_basis} (a). To learn a sparse basis for this image,
we decomposed the original image ($128 \times 128$ pixels) into all
possible $8\times8$ patches, which totals $14,641$ unique patches. The
patches were then reshaped into one-dimensional signals ${\vec y}_i$
each containing $64$ values. We chose $100$ basis functions (columns
of $\Phi$) to sparsely represent the input matrix
$Y$. Fig.~\ref{fi:cat_basis} (b) shows the $100$ basis functions that
were obtained by solving Eq.~\ref{eq:basis_learning_ch}. The
$100\times64$ matrix $\Phi$ was reshaped into $8\times8$ pixel basis
functions before plotting. Note that some of the basis functions
display complicated features, e.g., lines and ripples of different
widths and angles, whereas others are more uniform.

\begin{figure}
\includegraphics[width=3in]{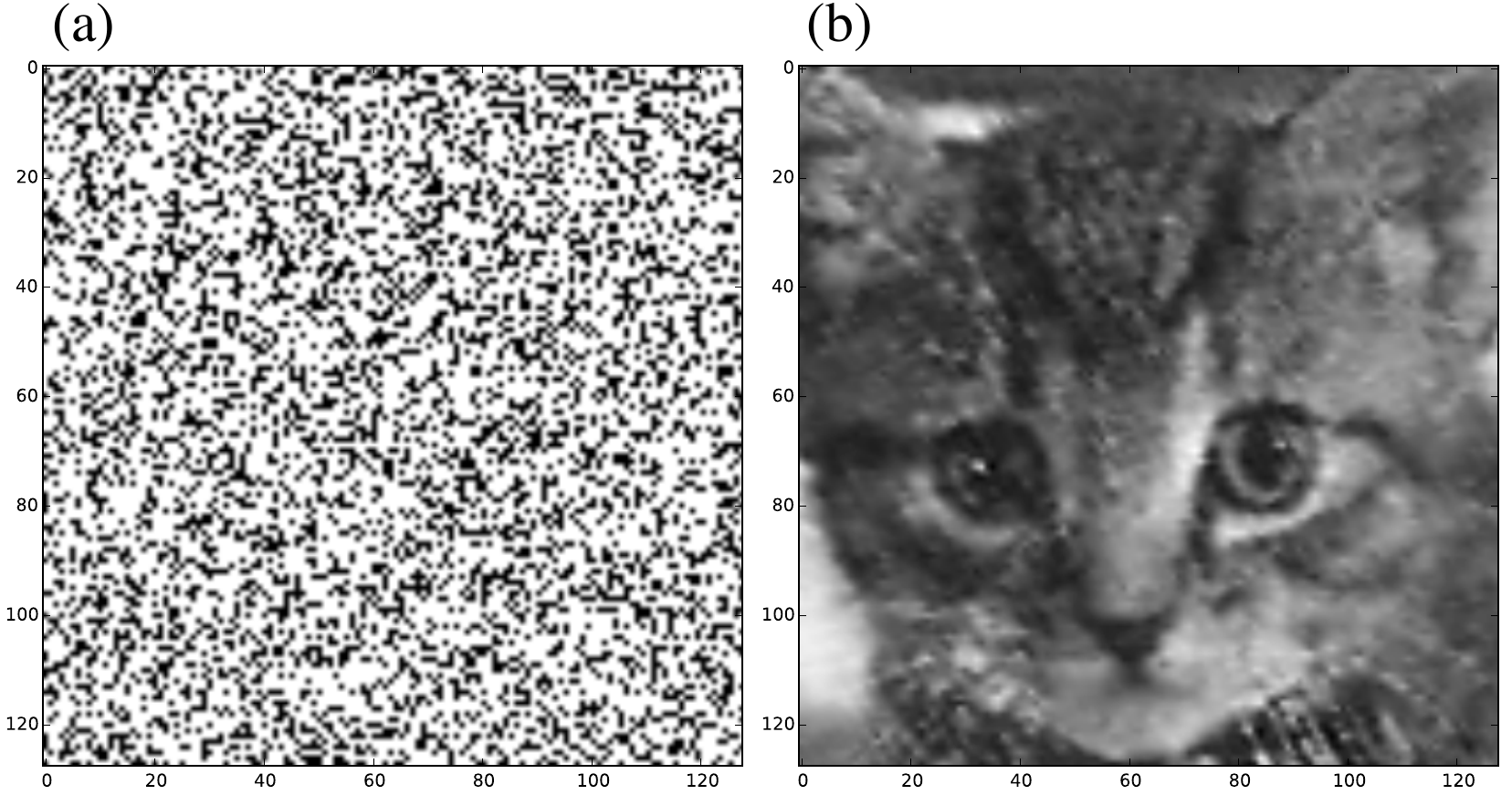}
\caption{(a) A random collection of pixels (sampling mask) that
represents $30\%$ of the image in Fig.~\ref{fi:cat_basis}~(a). (b)
Sparse basis learning reconstruction of the image using samples of
the image at the locations of the black pixels in (a).}
\label{fi:cat_rec}
\end{figure}

To demonstrate the utility of the sparse basis learning method, we seek to
recover the image in Fig.~\ref{fi:cat_basis} (a) from an undersampled
version using the learned basis functions in Fig.~\ref{fi:cat_basis}
(b) and then performing sparse reconstruction (Eq.~\ref{eq:L1min}) for
each $8\times8$ patch of the undersampled image.  For this example, we
randomly sampled $\approx 30$\% of the original image. In
Fig.~\ref{fi:cat_rec} (a), the black pixels indicate the random pixels
used for the sparse reconstruction of the undersampled 
image. We decompose the undersampled image into all
possible $8\times8$ patches, using only the measurements marked by
the black pixels in the sampling mask in Fig.~\ref{fi:cat_rec} (a). 
While the reconstruction of the image in Fig.~\ref{fi:cat_rec} (b) 
is somewhat grainy, this reconstruction method clearly resembles 
the original image even when it is $70$\% undersampled. 

In this work, we show that one may also use incomplete data to learn a
sparse basis. For example, the case of a discrete representation of a
two-dimensional system of ODEs is the same problem as basis learning
for image reconstruction (Fig.~\ref{fi:cat_basis}). However, learning
the basis from solutions of the system of ODEs, does not provide full
sampling of the signal ({\it i.e.} the right-hand side of the system
of ODEs in Eq.~\ref{eq:odesystem}), because the dynamics of the system
is strongly affected by the fixed point structure and the functions
are not uniformly sampled.

To learn a basis from incomplete data, we decompose the signal into
patches of a given size and then fill in the missing values with
random numbers.  We convert the padded patches ({\it i.e.} original
plus random signal) into a signal matrix $Y$ and learn a basis $\Phi$
to sparsely represent the signal by solving
Eq.~\ref{eq:basis_learning_ch}.  To recover the signal, we find a
sparse representation ${\hat c}$ of the unpadded signal ({\it i.e.}
without added random values) in the learned basis $\Phi$ by solving
Eq.~\ref{eq:underdeteqs}, where $P$ is the matrix that selects only
the signal entries that have been measured. When then obtain the
reconstructed patch by taking the product $\Phi {\hat c}$. We repeat
this process for each patch to reconstruct the full domain. For cases
in which we obtain different values for the signal at the same location from
different patches, we average the result.

\subsection{Models}
\label{models}

We test our methods for the reconstruction of systems of ODEs using
synthetic data, {\it i.e.} data generated by numerically solving
systems of ODEs, which allows us to test quantitatively the accuracy
as a function of the amount of data used in the
reconstruction. We present results from systems of ODEs in one, two,
and three dimensions with increasing complexity in the dynamics.  For
an ODE in one dimension (1D), we only need to reconstruct one
nonlinear function $f_1$ of one variable $x_1$. In two dimensions
(2D), we need to reconstruct two functions ($f_1$ and $f_2$) of two
variables ($x_1$ and $x_2$), and in three dimensions (3D), we need to
reconstruct three functions ($f_1$, $f_2$, and $f_3$) of three
variables ($x_1$, $x_2$, and $x_3$) to reproduce the dynamics of the
system. Each of the systems that we study possesses a different fixed
point structure in phase space. The 1D model has two stable fixed
points and one unstable fixed point, and thus all trajectories 
evolve toward one of the stable fixed points. The 2D model has one
saddle point and one oscillatory fixed point with closed orbits as
solutions. The 3D model we study has no stable fixed points and instead
possesses chaotic dynamics on a strange attractor. 

\paragraph{1D model} For 1D, we study the Reynolds model for the 
immune response to infection~\cite{day1}:
\begin{equation}
\frac{dx_1}{dt}= f_1(x_1)=k_{pg}x_1 \left(1-\frac{x_1}{x_\infty}\right)-\frac{k_{pm}
  s_m x_1}{\mu_m+k_{mp} x_1}\,,
\label{eq:mod1}
\end{equation}
where the pathogen load $x_1$ is unitless, and the other parameters $k_{pg}$, 
$k_{pm}$, $k_{mp}$, and $s_m$ have units of inverse hours.  
The right-hand side of Eq.~\ref{eq:mod1} is the sum of two terms. The
first term enables logistic growth of the pathogen load. In the absence
of any other terms, any positive initial value will cause $x_1$ to grow
logistically to the steady-state value $x_{\infty}$. The second term
mimics a local, non-specific response to an infection, which reduces 
the pathogen load. For small
values of $x_1$, the decrease is proportional to $x_1$. For larger values
of $x_1$, the decrease caused by the second term is constant.

\begin{figure}
\includegraphics[width=3.5in]{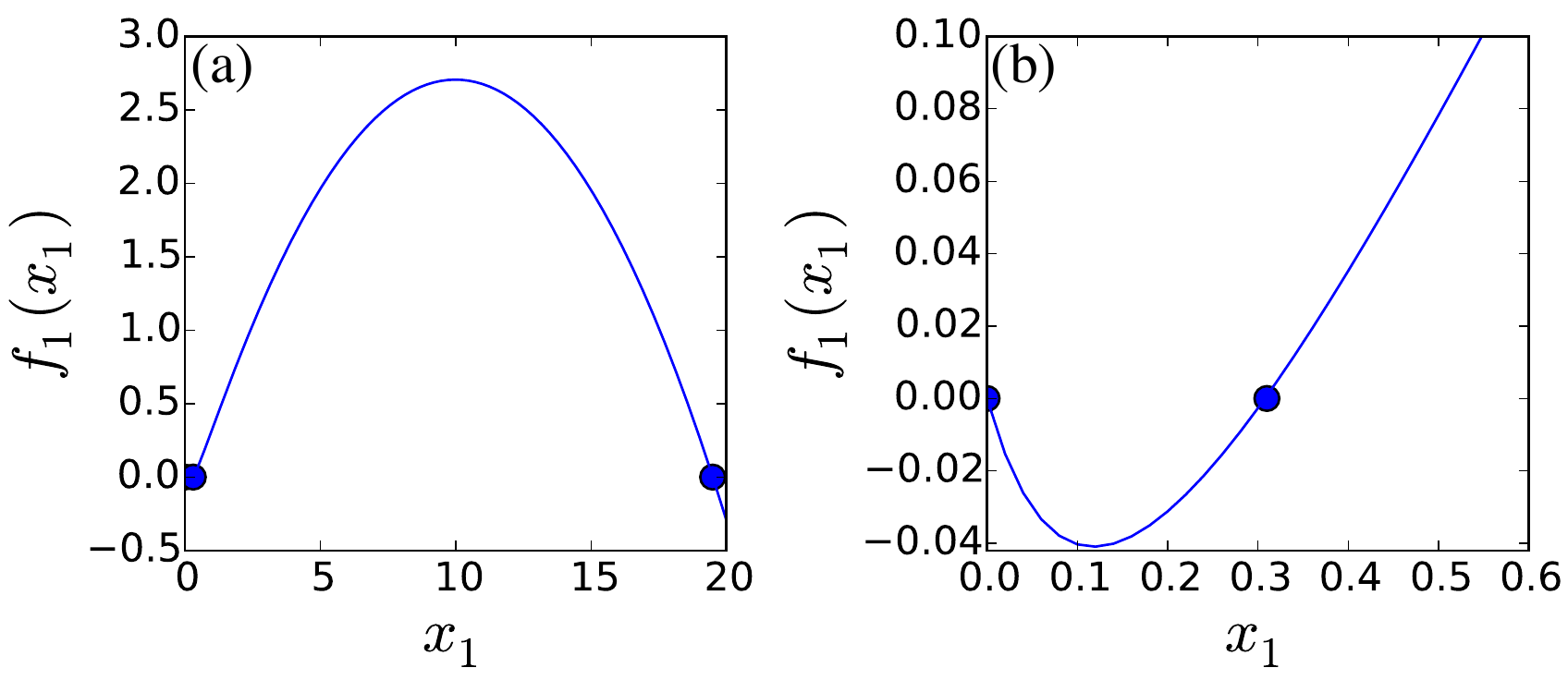}
\caption{(a) The function $f_1(x_1)$ for the
Reynolds ODE model in one spatial dimension (Eq.~\ref{eq:mod1}) for
the pathogen load in the range $0\leq x_1 \leq 20$. Fixed points are
marked by solid circles. (b) Close up of $f_1(x_1)$ near the stable
fixed point at $x_1=0$ and unstable fixed point at $x_1=0.31$. These
two fixed points are difficult to discern on the scale shown 
in panel (a). Trajectories with initial conditions between $x_1=0$ and $0.31$
will move towards the stable fixed point at $x_1=0$, while initial
conditions with $x_1>0.31$ will move toward the fixed point at
$x_1=19.49$. }
\label{fi:1d_model}
\end{figure}

\begin{figure}
\includegraphics[width=3.5in]{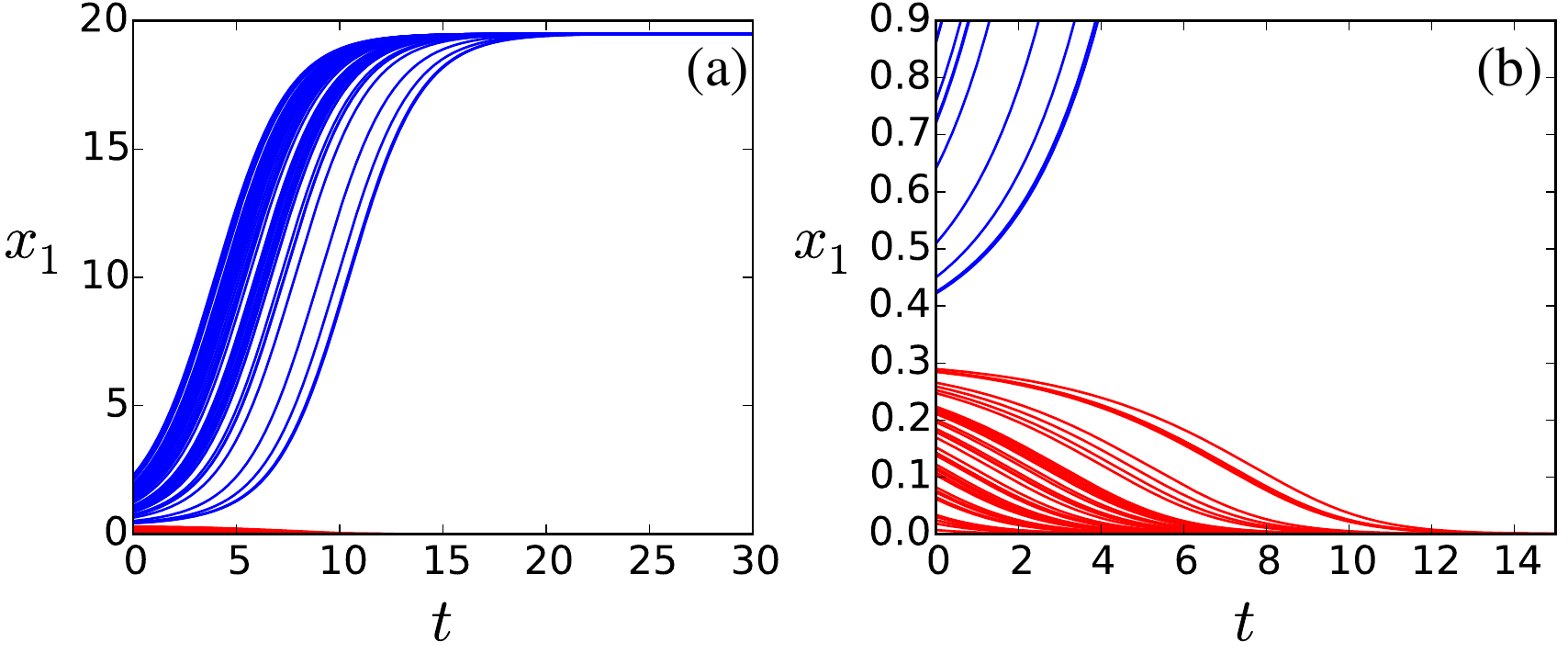}
\caption{(a) Pathogen load $x_1$ as a function of 
time $t$ in the range $0\leq t \leq 30$ for the Reynolds ODE model 
in one spatial dimension (Eq.~\ref{eq:mod1}).  (b) Close up of the 
region $0\leq t \leq 15$, which highlights the behavior of the system near 
the unstable fixed point at $x_1=0.31$. Solutions (red solid lines) with initial
conditions $0 \leq x_1 \leq 0.31$ are attracted to the stable fixed
point at $x_1=0$, while solutions (blue solid lines) with initial conditions 
$x_1>0.31$ are attracted to the stable fixed point at $x_1=19.49$.}
\label{fi:1d_traj}
\end{figure}

We employed the parameter values $k_{pg}=0.6$, $x_{\infty}=20$,
$k_{pm}=0.6$, $s_m=0.005$, $\mu_m=0.002$, and $k_{mp}=0.01$, which
were used in previous studies of this ODE model~\cite{day1,mai2015}. In this
parameter regime, Eq.~\ref{eq:mod1} exhibits two stable fixed points
at $x_1=0$ and $19.49$ and one unstable fixed point, separating the two
stable fixed points, at $x_1=0.31$ (Fig.~\ref{fi:1d_model}). As shown in
Fig.~\ref{fi:1d_traj}, solutions to Eq.~\ref{eq:mod1} with initial
conditions $0 \leq x_1 \leq 0.31$ are attracted to the stable fixed
point at $x_1=0$, while solutions with initial conditions $x_1 > 0.31$ are
attracted to the stable fixed point at $x_1=19.49$.

\paragraph{2D model} In 2D, we focused on the Lotka-Volterra system 
of ODEs that describe predator-prey dynamics~\cite{murray2002}:
\begin{equation}
\label{eq:mod2}
\begin{aligned}
\frac{dx_1}{dt}&= f_1(x_1,x_2) = \alpha \,x_1 - \beta\, x_1 x_2\\ \frac{dx_2}{dt}&= f_2(x_1,x_2)= -
\gamma\, x_2+\delta\, x_1 x_2\,,
\end{aligned}
\end{equation}
where $x_1$ and $x_2$ describe the prey and predator population sizes,
respectively, and are unitless. In this model, prey have a natural
growth rate $\alpha$. In the absence of predators, the prey population
$x_1$ would grow exponentially with time. With predators present, the
prey population decreases at a rate proportional to the product of
both the predator and prey populations with a proportionality constant
$\beta$ (with units of inverse time). Without predation, the predator
population $x_2$ would decrease at death rate $\gamma$. With the
presence of prey $x_1$, the predator population grows proportional to
the product of the two population sizes $x_1$ and $x_2$ with a
proportionality constant $\delta$ (with units of inverse
time). 

For the Lotka-Volterra system of ODEs, there are two fixed points, one
at $x_1=0$ and $x_2=0$ and one at $x_1=\gamma/\delta$ and
$x_2=\alpha/\beta$. The stability of the fixed points is determined by
the eigenvalues of the Jacobian matrix evaluated at the fixed points. The
Jacobian of the Lotka-Volterra system is given by
\begin{equation}
J_{LV} = 
\begin{pmatrix}
\alpha - \beta x_2 & -\beta x_1\\
\delta x_2 & -\gamma +\delta x_1
\end{pmatrix}\,.	
\label{eq:J_LV}
\end{equation}
The eigenvalues of the Jacobian $J_{LV}$ at the
origin are ${\cal J}_1 = \alpha$, ${\cal J}_2 = -\gamma$. Since the
model is restricted to positive parameters, the fixed point at the
origin is a saddle point. The interpretation is that for small
populations of predator and prey, the predator population decreases
exponentially due to the lack of a food source. While unharmed by the
predator, the prey population can grow exponentially, which drives the
system away from the zero population state, $x_1=0$ and $x_2=0$.

\begin{figure}
\includegraphics[width=3in]{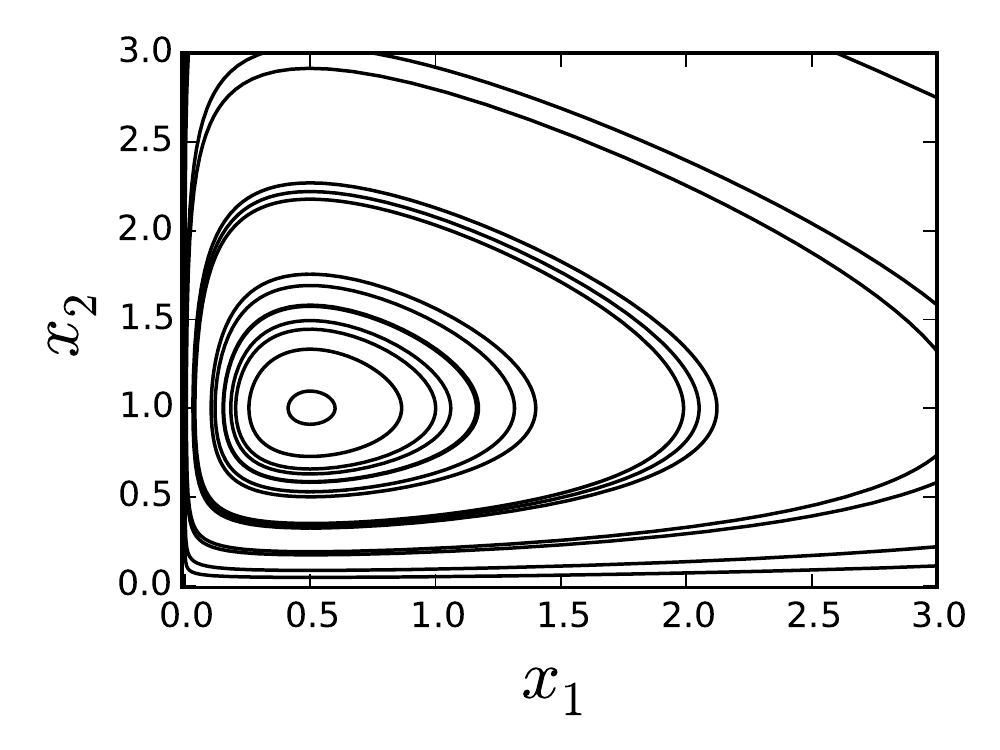}
\caption{Solutions of the
two-dimensional Lotka-Volterra ODE model (Eq.~\ref{eq:mod2}) with
$\alpha=0.4$, $\beta=0.4$, $\gamma=0.1$, and $\delta=0.2$ plotted
parametrically ($x_2(t)$ versus $x_1(t)$) for $15$ random initial
conditions in the domain $0\leq x_1 \leq 0.5$ and $0\leq x_2 \leq
1$. This model has an unstable fixed point at $x_1=0$ and $x_2=0$
and an oscillatory fixed point at $x_1 = 0.5$ and $x_2=1$.}
\label{fi:2d_traj}
\end{figure}

The eigenvalues of the Jacobian $J_{LV}$ at the second fixed point
$x_1=\gamma/\delta$ and $x_2=\alpha/\beta$ are purely imaginary
complex conjugates, ${\cal J}_1 = -i \sqrt{\alpha\gamma}$ and ${\cal
  J}_2 = i \sqrt{\alpha\gamma}$, where $i^2=-1$. The purely imaginary
fixed point causes trajectories to revolve around it and form closed
orbits. The interpretation of this fixed point is that the predators
decrease the number of prey, then the predators begin to die due to a
lack of food, which in turn allows the prey population to grow. The
growing prey population provides an abundant food supply for the
predator, which allows the predator to grow faster than the food
supply can sustain.  The prey population then decreases and the cycle
repeats. For the results below, we chose the parameters $\alpha=0.4$,
$\beta=0.4$, $\gamma=0.1$, and $\delta=0.2$ for the Lotka-Volterra
system, which locates the oscillatory fixed point at $x_1=0.5$ and
$x_2=1.0$ (Fig. \ref{fi:2d_traj}).

\paragraph{3D model} In 3D, we focused on the Lorenz system of ODEs~\cite{lorenz1963}, which describes fluid motion in a container that is heated from 
below and cooled from above: 
\begin{equation}
\label{eq:mod3}
\begin{aligned}
\frac{dx_1}{dt}&=f_1(x_1,x_2,x_3)=\sigma(x_2 - x_1 )\\ \frac{dx_2}{dt}&=f_2(x_1,x_2,x_3)=x_1(\rho - x_3)
- x_2\\ \frac{dx_3}{dt}&=f_3(x_1,x_2,x_3)=x_1 x_2 - \beta x_3\,,
\end{aligned}
\end{equation}
where $\sigma, \rho, \beta$ are positive, dimensionless parameters
that represent properties of the fluid. In different parameter
regimes, the fluid can display quiescent, convective, and chaotic
dynamics. The three dimensionless variables $x_1$, $x_2$, and $x_3$
describe the intensity of the convective flow, temperature difference
between the ascending and descending fluid, and spatial dependence of the
temperature profile, respectively.

\begin{figure}
\includegraphics[width=3.5in]{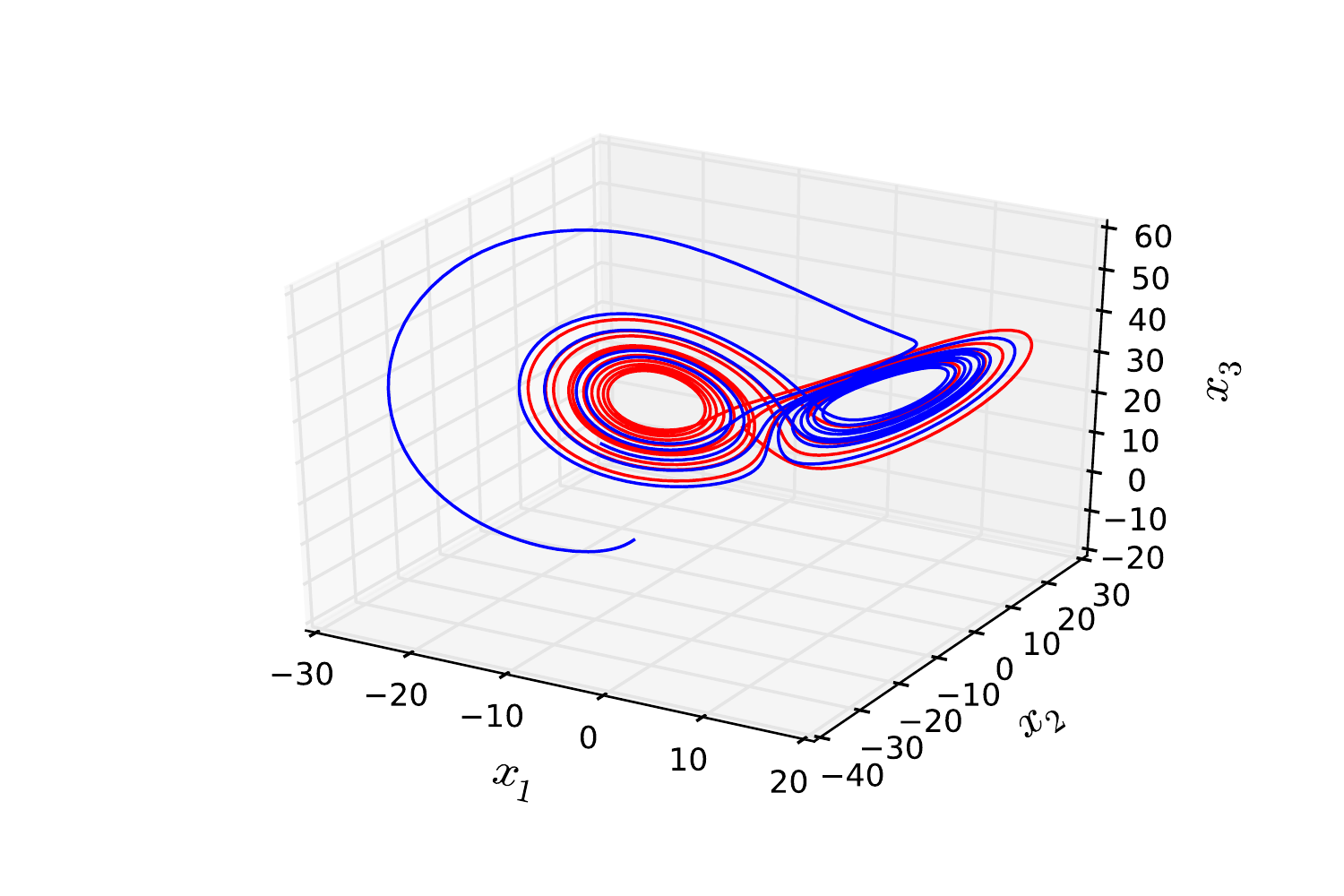}
\caption{Solutions of the
three-dimensional Lorenz ODE model (Eq.~\ref{eq:mod3}) with $\sigma =10$, 
$\rho=28$, and $\beta=8/3$ plotted
parametrically in the $x_1$-$x_2$-$x_3$ plane for two initial conditions 
$(x_1,x_2,x_3)=(-7.57,11.36,9.51)$ (red solid line) and $(-12.17,-5.23,
-11.60)$ (blue solid line). For this parameter regime, the system has three unstable fixed 
points, lives on the Lorenz attractor, and possesses chaotic dynamics.}
\label{fi:3d_traj}
\end{figure}

The system possesses three fixed points at
$(x_1,x_2,x_3)=(0,0,0)$,
$(-\beta^{1/2}(\rho-1)^{1/2},-\beta^{1/2}(\rho-1)^{1/2},\rho-1)$,
and
$(\beta^{1/2}(\rho-1)^{1/2},\beta^{1/2}(\rho-1)^{1/2},\rho-1)$. The
Jacobian of the system is given by
\begin{equation}
J_{L} = 
\begin{pmatrix}
- \sigma & \sigma & 0\\
\rho - x_3 & -1 & -x_1\\
x_2 & x_1 & -\beta
\end{pmatrix}\,.	
\label{eq:J_L}
\end{equation}
When we evaluate the Jacobian (Eq.~\ref{eq:J_L}) at the fixed points,
we find that each of the three eigenvalues possesses two stable and
one unstable eigendirection in the parameter regime $\sigma = 10$,
$\rho = 28$ and $\beta = 8/3$.  With these parameters, the Lorenz 
system displays chaotic dynamics
with Lyapunov exponents $\ell_1=0.91$, $\ell_2=0.0$, and
$\ell_3=-14.57$. In Fig.~\ref{fi:3d_traj}, we show the time
evolution of two initial conditions in $x_1$-$x_2$-$x_3$  configuration space
for this parameter regime.

\section{Results}
\label{ofd_Results}

In this section, we present the results of our methodology for ODE
reconstruction of data generated from the three systems of ODEs
described in Sec.~\ref{models}.  For each system, we measure the
accuracy of the reconstruction as a function of the size of the
sections used to decompose the signal for basis learning, the sampling
time interval between time series measurements, and the number of
trajectories. For each model, we make sure that the total integration
time is sufficiently large that the system can reach the stable
fixed points or sample the chaotic attractor in the case of the Lorenz system.

\subsection{Reconstruction of ODEs in 1D}

We first focus on the reconstruction of the Reynolds ODE model in 1D
(Eq.~\ref{eq:mod1}) using time series data.  We discretized the domain
$0 \leq x_1 \leq 19.5$ using $256$ points, $i=0,\ldots,255$.  Because
the unstable fixed point at $x_1 = 0.31$ is much closer to the stable
fixed point at $x_1 = 0$ than to the stable fixed point at $x_1 =
19.49$, we sampled more frequently in the region $0 \leq x_1 \leq 0.6$
compared to the region $0.6 < x_1 \leq 19.5$.  In particular, we
uniformly sampled $128$ points from the small domain, and uniformly
sampled the same number of points from the large domain.

\begin{figure}[!ht]
\begin{center}
\includegraphics[width=3in]{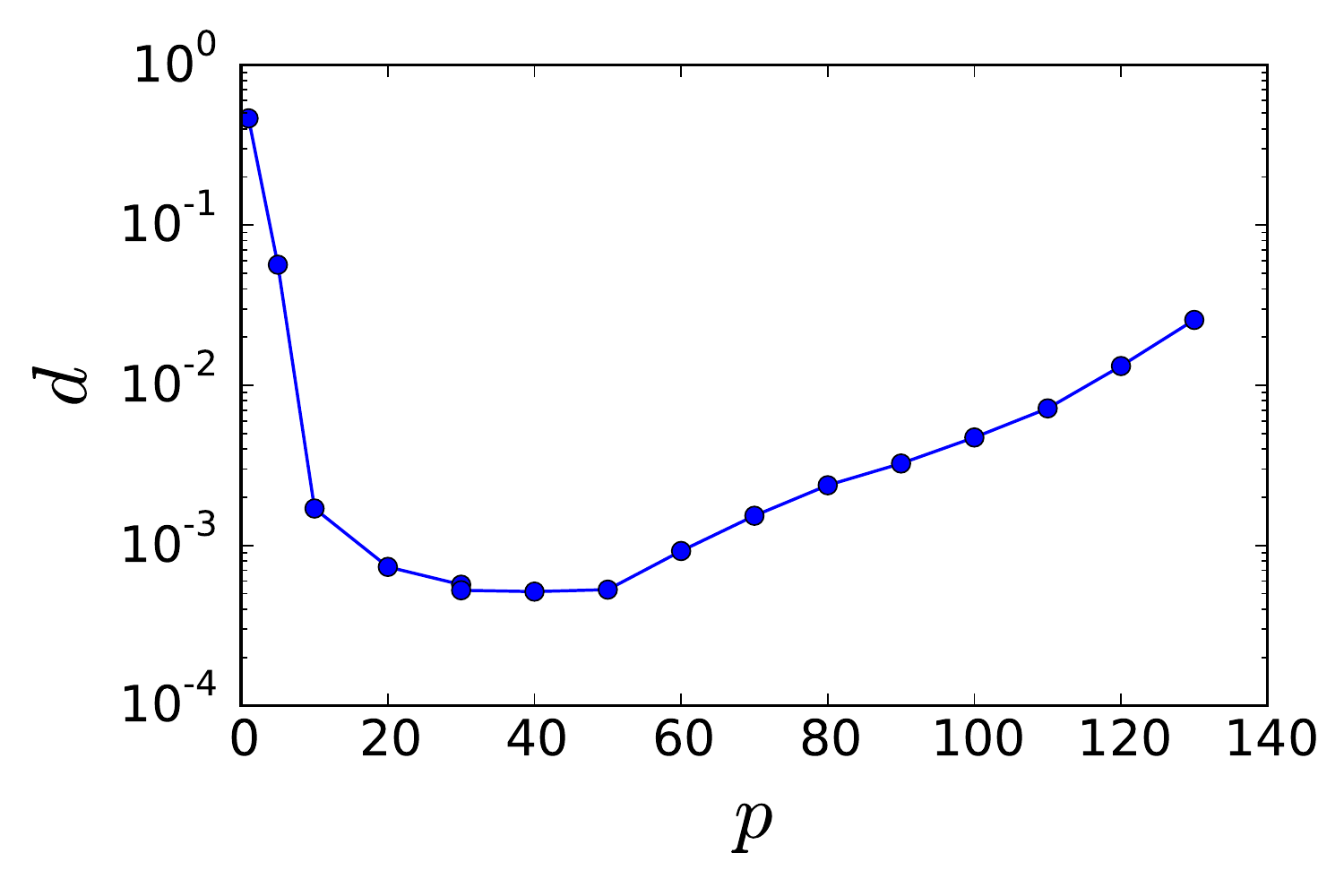}
\end{center}
\caption{Error $d$ in the ODE reconstruction for the 1D Reynolds model
(Eq.~\ref{eq:mod1}) as a function of the patch size $p$ used for
basis learning using $N_{t}=10$ trajectories with 
sampling time interval $\Delta t=1$.}
\label{fi:1d_err_vs_b_size}
\end{figure}

In Fig.~\ref{fi:1d_err_vs_b_size}, we show the error $d$
(Eq.~\ref{eq:distance}) in recovering the right-hand side of
Eq.~\ref{eq:mod1} ($f_1(x_1)$) as a function of the size $p$ of the
patches used for basis learning. Each data point in
Fig.~\ref{fi:1d_err_vs_b_size} represents an average over $20$
reconstructions using $N_t=10$ trajectories with a sampling time
interval $\Delta t = 1$.  We find that the error $d$ achieves a
minimum below $10^{-3}$ in the patch size range $30 < p < 50$. Basis
sizes that are too small do not adequately sample $f_1(x_1)$, while
basis patches that are too large do not include enough variability to
select a sufficiently diverse basis set to reconstruct $f_1(x_1)$. For
example, in the extreme case that the basis patch size is the same
size as the signal, we are only able to learn the input data itself,
which may be missing data. For the remaining studies of the 1D model,
we set $p=50$ as the basis patch size.

\begin{figure}[!ht]
\begin{center}
\includegraphics[width=3in]{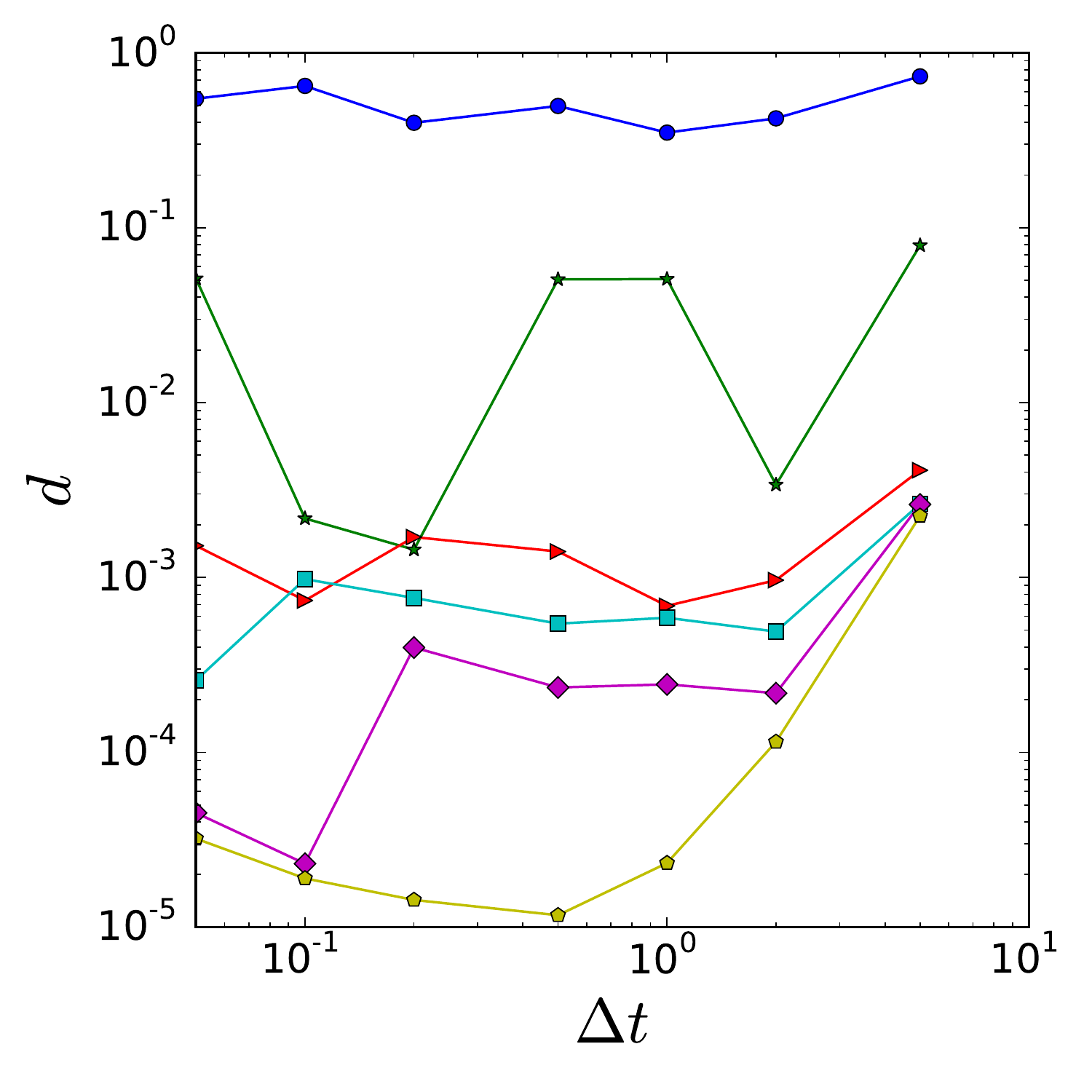}
\end{center}
\caption{Reconstruction error $d$ for the 1D Reynolds model as a
function of the sampling interval $\Delta t$ for several different
numbers of trajectories $N_t=1$ (circles), $5$ (stars), $10$
(triangles), $20$ (squares), $50$ (diamonds), and $200$ (pentagons) used in the reconstruction. The data for each $N_t$ is averaged over $20$
independent reconstructions.}
\label{fi:1d_err_vs_dt}
\end{figure}

\begin{figure}[!ht]
\begin{center}
\includegraphics[width=3in]{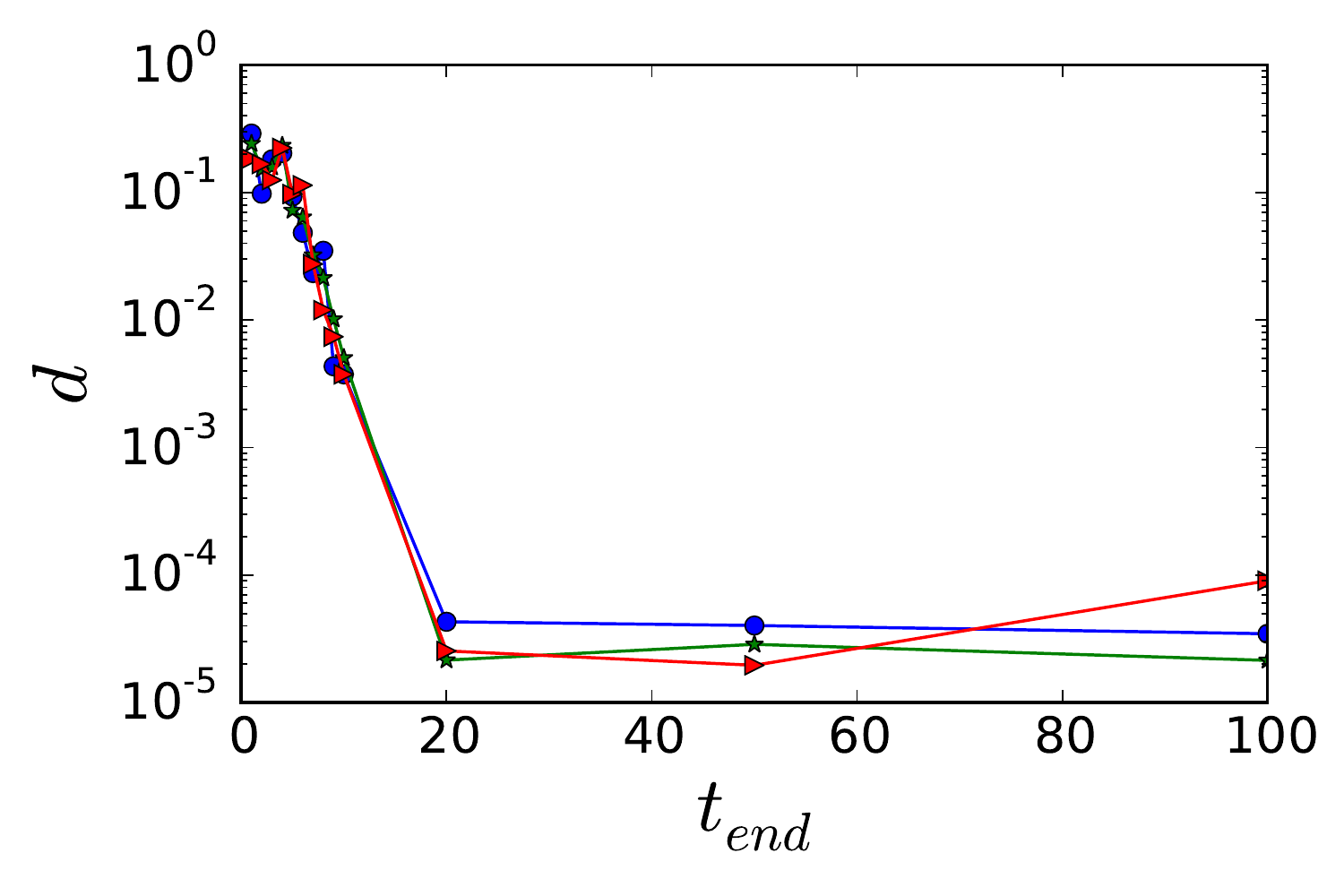}
\end{center}
\vspace{-0.2in}
\caption{Reconstruction error $d$ for the 1D Reynolds ODE model as a function
of the total integration time $t_{end}$ used for the reconstruction for
$N_t=20$ trajectories at several values of the sampling time $\Delta t=0.05$
(circles), $0.1$ (stars), and $0.5$ (triangles).}
\label{fi:1d_err_vs_tend}
\end{figure}

In Fig.~\ref{fi:1d_err_vs_dt}, we plot the error in the reconstruction
of $f_1(x_1)$ as a function of the sampling time interval $\Delta t$ for
several numbers of trajectories $N_{t}=1$, $2$, $20$, $50$, and
$200$. We find that the error decreases with the number of
trajectories used in the reconstruction. For $N_t = 1$, the error is
large with $d > 0.1$. For large numbers of trajectories ({\it e.g.}
$N_t=200$), the error decreases with decreasing $\Delta t$, reaching
$d \sim 10^{-5}$ for small $\Delta t$. The fact that the error in the
ODE reconstruction increases with $\Delta t$ is consistent with notion
that the accuracy of the numerical derivative of each trajectory
decreases with increasing sampling interval. In
Fig.~\ref{fi:1d_err_vs_tend}, we show the error in the reconstruction
of $f_1(x_1)$ as a function of the total integration time $t_{end}$.  We
find that $d$ decreases strongly as $t_{end}$ increases for $t_{end} <
20$. For $t_{end} > 20$, $d$ reaches a plateau value below $10^{-4}$,
which depends weakly on $\Delta t$.  For characteristic time scales $t
> 20$, the Reynolds ODE model reaches one of the two stable fixed
points, and therefore $d$ becomes independent of $t_{end}$.

\begin{figure}[!ht]
\begin{center}
\includegraphics[width=3in]{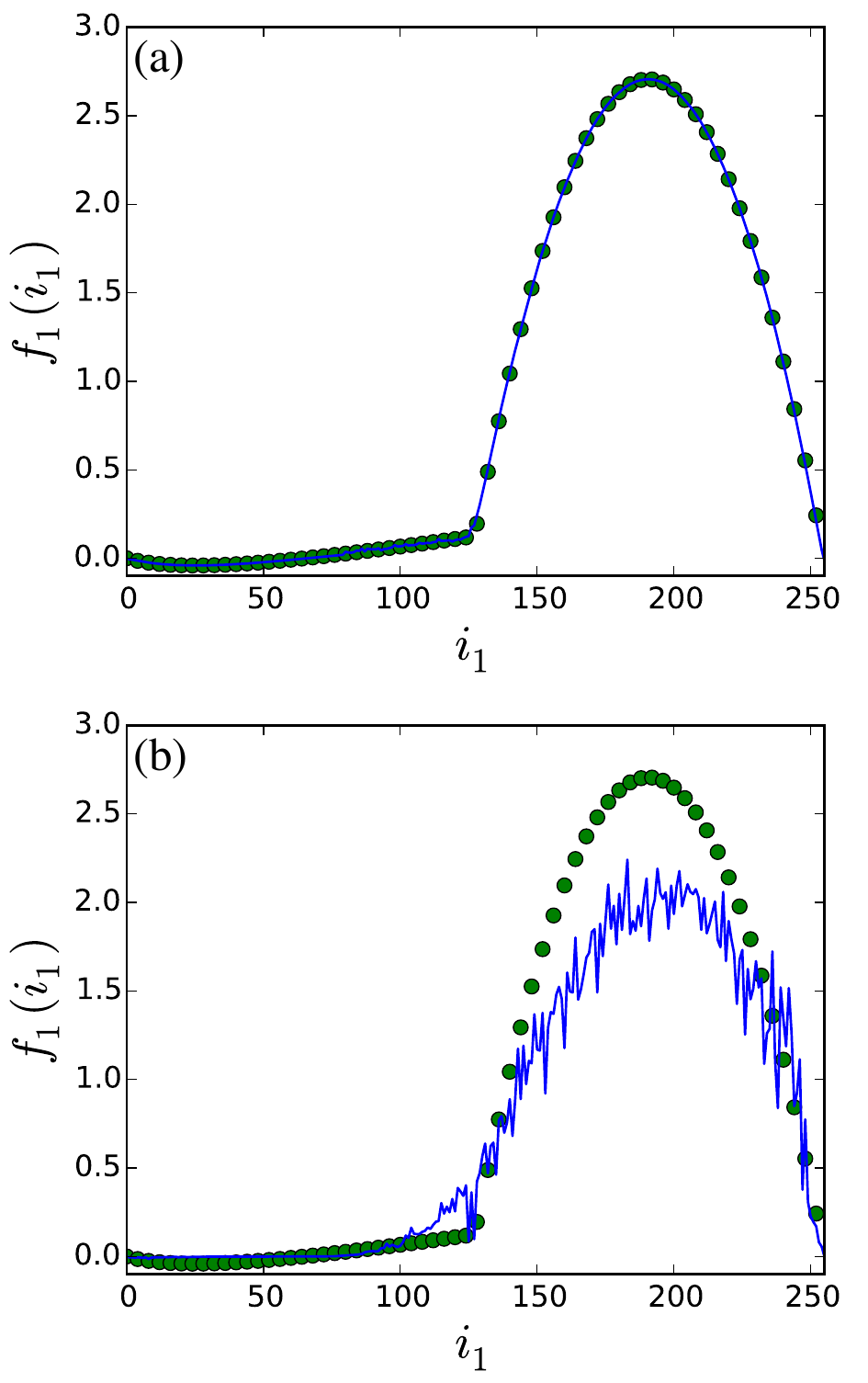}
\end{center}
\caption{Reconstructions (solid blue lines) of $f_1(i_1)$ for the 1D
Reynolds ODE model in Eq.~\ref{eq:mod1} using (a) $N_{t}=50$ and $\Delta
t=0.1$ and (b) $N_{t}=10$ and $\Delta t=5$. $f_1$ is discretized using $256$ points. The 
indices $i_1=0,\ldots,127$ indicate uniformly 
spaced $x_1$ values in the interval $0 \leq x_1 \leq 0.6$, 
and $i_1=128,\ldots,256$ indicate uniformly spaced $x_1$ values
in the interval $0.6 < x_1 \leq 19.5$. The exact expression for 
$f_1(i_1)$ is represented by the green circles.}
\label{fi:1d_example_rec_gb}
\end{figure}

In Fig.~\ref{fi:1d_example_rec_gb}, we compare accurate (using
$N_{t}=50$ and $\Delta t=0.1$) and inaccurate (using $N_{t}=10$ and
$\Delta t=5$) reconstructions of $f_1(i_1)$ for the 1D Reynolds ODE
model.  Note that we plot $f_1$ as a function of the scaled variable
$i_1$. The indexes $i_1=0,\ldots,127$ indicate uniformly spaced $x_1$
values in the interval $0 \leq x_1 \leq 0.6$, and $i_1=128,\ldots,256$
indicate uniformly spaced $x_1$ values in the interval $0.6 < x_1 \leq
19.5$.

We find that using large $\Delta t$ gives rise to inaccurate
measurements of the time derivative of $x_1$ and, thus of
$f_1(x_1)$. In addition, large $\Delta t$ does not allow dense
sampling of phase space, especially in regions where the trajectories
evolve rapidly. The inaccurate reconstruction in
Fig.~\ref{fi:1d_example_rec_gb} (b) is even worse than it seems at
first glance. The reconstructed function is identically zero over a
wide range of $i_1$ ($0 \leq i_1 \leq 50$) where $f(i_1)$ is not well
sampled, since the default output of a failed reconstruction is
zero. It is a coincidence that $f_1(i_1)\sim 0$ in Eq.~\ref{eq:mod1}
over the same range of $i_1$.

\begin{figure}[!ht]
\begin{center}
\includegraphics[width=3.2in]{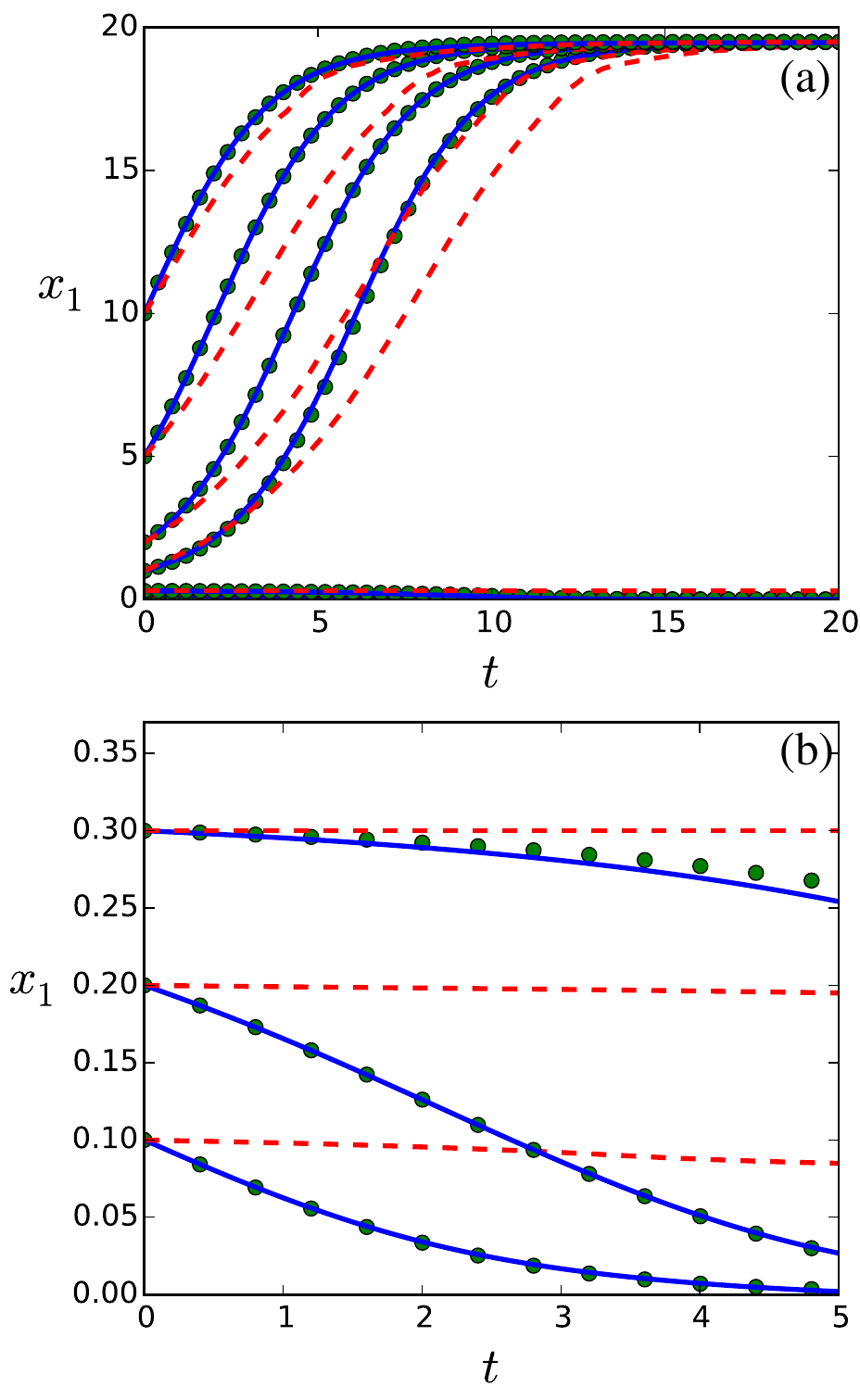}
\end{center}
\caption{(a) Comparison of the trajectories $x_1(t)$ for the 1D
Reynolds ODE model using the accurate (solid blue lines) and
inaccurate (dashed red lines) reconstructions of $f_1(x_1)$ shown in
Fig.~\ref{fi:1d_example_rec_gb} (a) and (b), respectively, for six 
initial conditions: $x_1(0)= 0.1$, $0.2$, $0.3$, $1$, $2$, $5$, and 
$10$. The trajectories obtained from $f_1(x_1)$ in Eq.~\ref{eq:mod1} are 
given by the open red circles.  (b) Close up of the trajectories in the
domain $0 \leq x_1 \leq 0.4$.}
\label{fi:1d_traj_rec_gb_original}
\end{figure}

We now numerically solve the reconstructed 1D Reynolds ODE model for
different initial conditions and times comparable to $t_{end}$ and
compare these trajectories to those obtained from the original model
(Eq.~\ref{eq:mod1}). In Fig.~\ref{fi:1d_traj_rec_gb_original}, we
compare the trajectories $x_1(t)$ for the accurate ($d\sim 10^{-5}$;
Fig.~\ref{fi:1d_example_rec_gb} (a)) and inaccurate ($d \sim 10^{-2}$;
Fig.~\ref{fi:1d_example_rec_gb} (b)) representations of $f_1(x_1)$ to
the original model for six initial conditions. All of the trajectories
for the accurate representation of $f_1(x_1)$ are nearly
indistinguishable from the trajectories for the original model,
whereas we find large deviations between the original and
reconstructed trajectories even at short times for the inaccurate
representation of $f_1(x_1)$.

\subsection{Reconstruction of Systems of ODEs in 2D}

We now investigate the reconstruction accuracy of our method for the
Lotka-Volterra system of ODEs in 2D.  We find that the results are
qualitatively similar to those for the 1D Reynolds ODE model.  We map
the numerical derivatives for $N_t$ trajectories with a sampling time
interval $\Delta t$ onto a $128 \times 128$ grid with $0 \leq x_1, x_2
\leq 2$. Similar to the 1D model, we find that the error in the
reconstruction of $f_1(x_1,x_2)$ and $f_2(x_1,x_2)$ possesses a
minimum as a function of the patch area used for basis learning, where
the location and value at the minimum depends on the parameters used for
the reconstruction.  For example, for $N_t = 200$, $\Delta t=0.1$, and
averages over $20$ independent runs, the error reaches a minimum ($d
\approx 10^{-5}$) near $p^2_{\rm min} \approx 100$.

\begin{figure}[!ht]
\begin{center}
\includegraphics[width=3in]{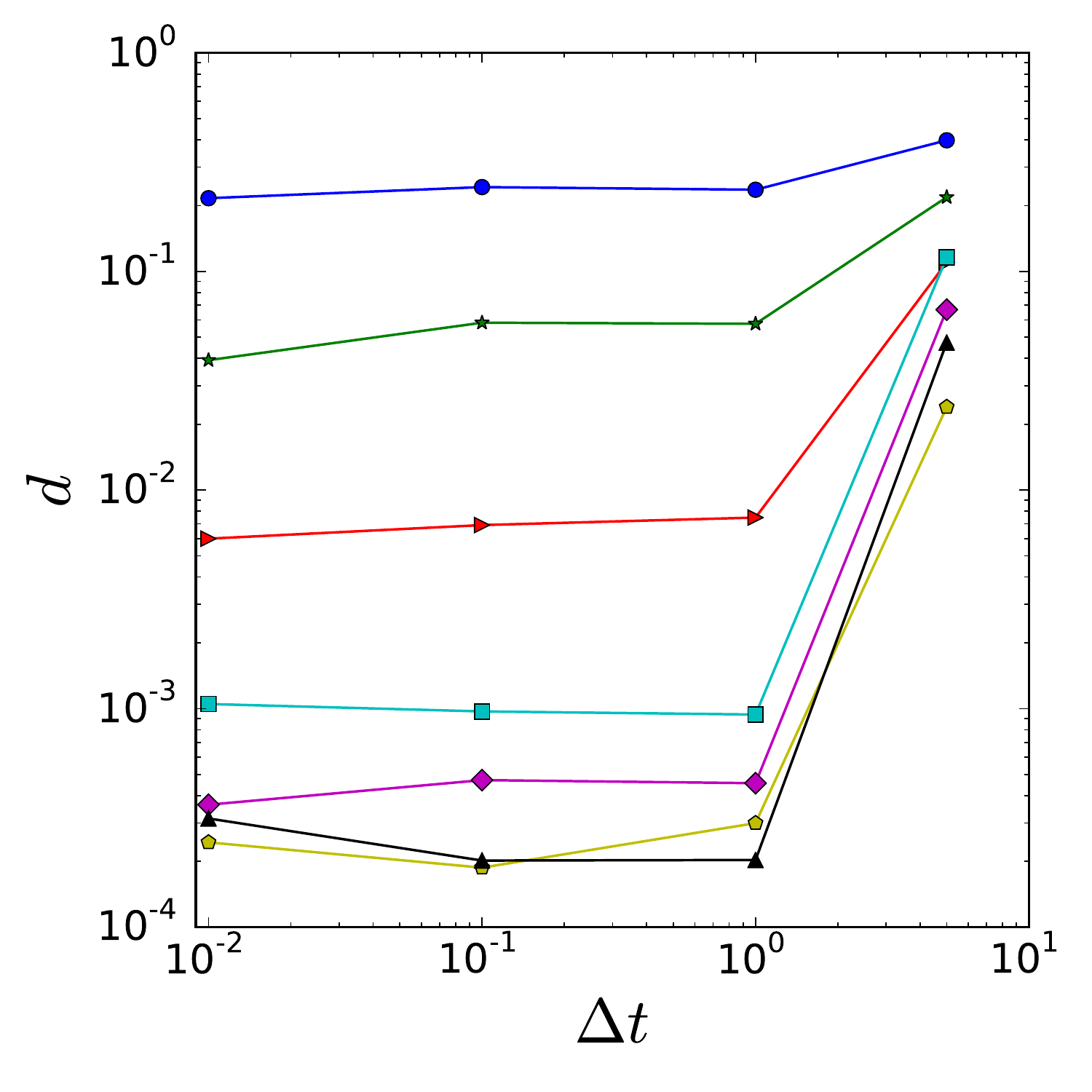}
\end{center}
\caption{Reconstruction error $d$ for the 2D Lotka-Volterra
model (Eq.~\ref{eq:mod2}) as a function of the sampling time interval 
$\Delta t$ for different numbers of trajectories $N_t=5$ (circles), $10$ (stars), $20$ (rightward triangles), $50$ (squares), $100$ (diamonds), $200$ (pentagons), and $500$ (upward triangles) averaged over $20$ independent 
runs.}
\label{fi:2d_err_vs_dt} 
\end{figure}

In Fig. \ref{fi:2d_err_vs_dt}, we show the reconstruction error $d$ as
a function of the sampling time interval $\Delta t$ for several values
of $N_{t}$ from $5$ to $500$ trajectories, for a total time $t_{end}$ that 
allows several revolutions around the closed orbits, and for patch size 
$p^2_{\rm min}$. As in 1D, we find that
increasing $N_t$ reduces the reconstruction error. For $N_{t}=5$, $d
\sim 10^{-1}$, while $d < 10^{-3}$ for $N_{t}=500$. $d$ also decreases
with decreasing $\Delta t$, although $d$ reaches a plateau in the small
$\Delta t$ limit, which depends on the number of trajectories included
in the reconstruction. 

\begin{figure}[!ht]
\begin{center}
\includegraphics[width=3.4in]{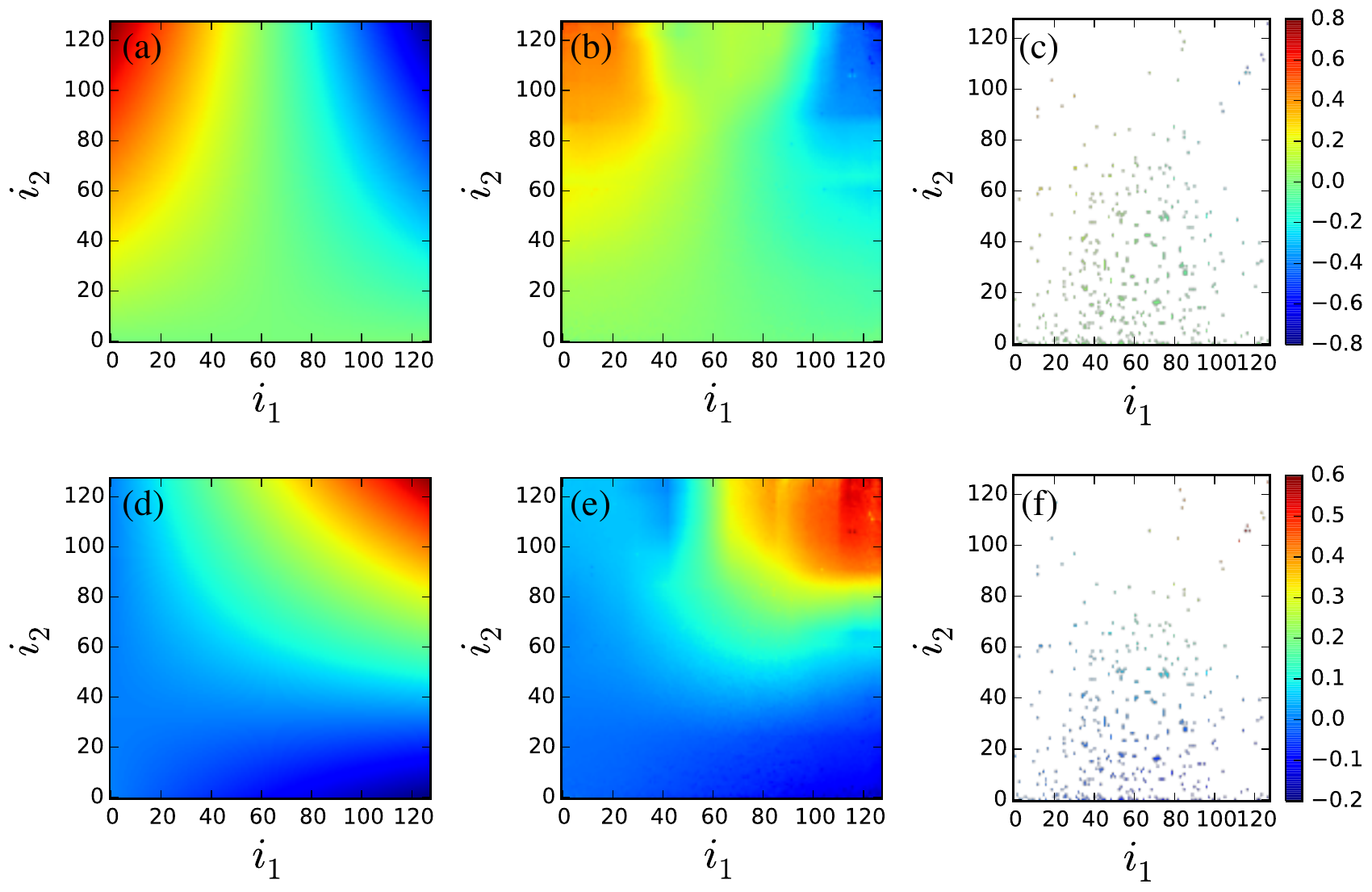}
\end{center}
\caption{Example of inaccurate reconstructions of $f_1(i_1,i_2)$ and
$f_2(i_1,i_2)$ (with errors $d=0.04$ and $0.03$, respectively) for
the 2D Lotka-Volterra system of ODEs (\ref{eq:mod2}). The indexes 
$i_{1,2}=0,\ldots,127$ indicate uniformly spaced $x_1$ and 
$x_2$ values in the interval $0 \leq x_{1,2} \leq 2$. Panels (a) and
(d) give the exact functions $f_1(i_1,i_2)$ and $f_2(i_1,i_2)$, panels (b)
and (e) give the reconstructions, and panels (c) and (f) indicate
the $i_1$-$i_2$ values that were sampled (2\% of the $128\times128$ grid). The white regions in panel (c) and (f) indicate missing data. In the top row, the color scale varies from $-0.8$ to $0.8$
(from dark blue to red).  In the bottom row, the color scale varies
from $-0.2$ to $0.6$ (from dark blue to red).  This reconstruction
was obtained using $N_{t} = 30$ trajectories, a sampling interval
$\Delta t=5$, and a basis patch area $p^2=625$.}
\label{fi:2d_example_rec_b}
\end{figure}

\begin{figure}[!ht]
\begin{center}
\includegraphics[width=3.4in]{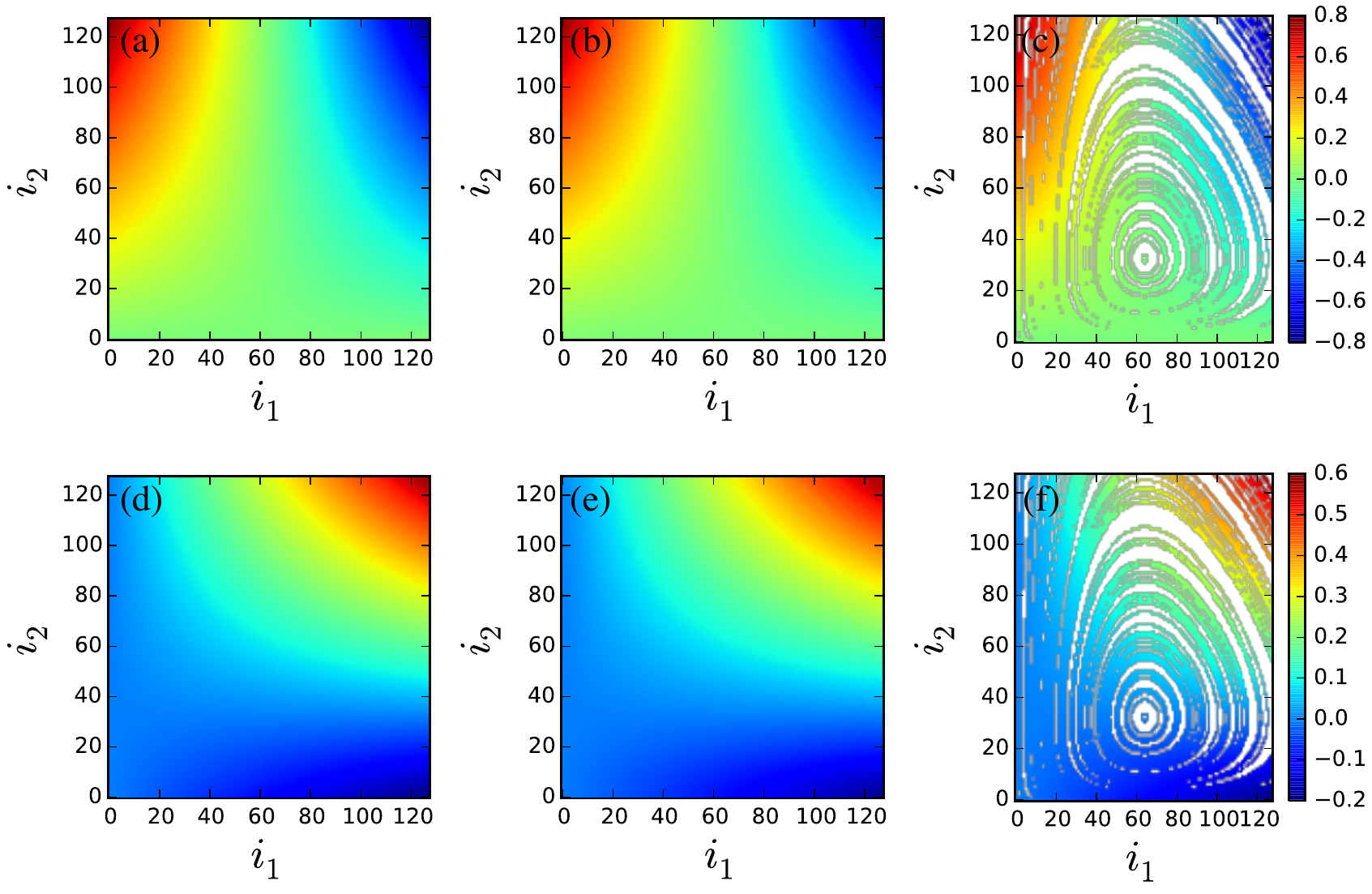}
\end{center}
\caption{Example of accurate reconstructions of $f_1(i_1,i_2)$ and
$f_2(i_1,i_2)$ (with errors $d=4 \times 10^{-5}$ and $5 \times 
10^{-5}$, respectively) for the 2D Lotka-Volterra system 
of ODEs (\ref{eq:mod2}). The indexes $i_{1,2}=0,\ldots,128$ indicate 
uniformly spaced $x_1$ and $x_2$ values in the interval $0 \leq x_{1,2} \leq
2$. Panels (a) and (d) give the exact functions 
$f_1(i_1,i_2)$ and $f_2(i_1,i_2)$, panels (b) and (e) give the 
reconstructions, and panels (c) and (f) indicate the $i_1$-$i_2$ 
values that were sampled ($68\%$ of the $128\times 128$ grid). The white regions in panel (c) and (f) indicate missing data. The color 
scales 
are the same as in Fig.~\ref{fi:2d_example_rec_b}. This reconstruction 
was obtained using $N_{t} = 100$
  trajectories, a sampling interval of $\Delta t=0.01$, and a basis
patch area $p^2=625$.}
\label{fi:2d_example_rec_g}
\end{figure}

In Figs.~\ref{fi:2d_example_rec_b} and~\ref{fi:2d_example_rec_g}, we
show examples of inaccurate ($d \sim 10^{-2}$) and accurate ($d \sim
10^{-5}$) reconstructions of $f_1(i_1,i_2)$ and $f_2(i_1,i_2)$.  The
indexes $i_{1,2}=0,\ldots,127$ indicate uniformly spaced $x_1$ and
$x_2$ values in the interval $0 \leq x_{1,2} \leq 2$. The parameters
for the inaccurate reconstructions were $N_t=30$ trajectories and
$\Delta t = 5$ (enabling $f_1$ and $f_2$ to be sampled only over $2\%$
of the domain), whereas the parameters for the accurate
reconstructions were $N_t = 100$ trajectories and $\Delta t = 0.01$
(enabling $f_1$ and $f_2$ to be sampled over $68\%$ of the
domain). These results emphasize that even though the derivatives are
undetermined over nearly one-third of the domain, we can reconstruct the
functions $f_1$ and $f_2$ extremely accurately over the full domain.

\begin{figure}[!ht]
\begin{center}
\includegraphics[width=3.5in]{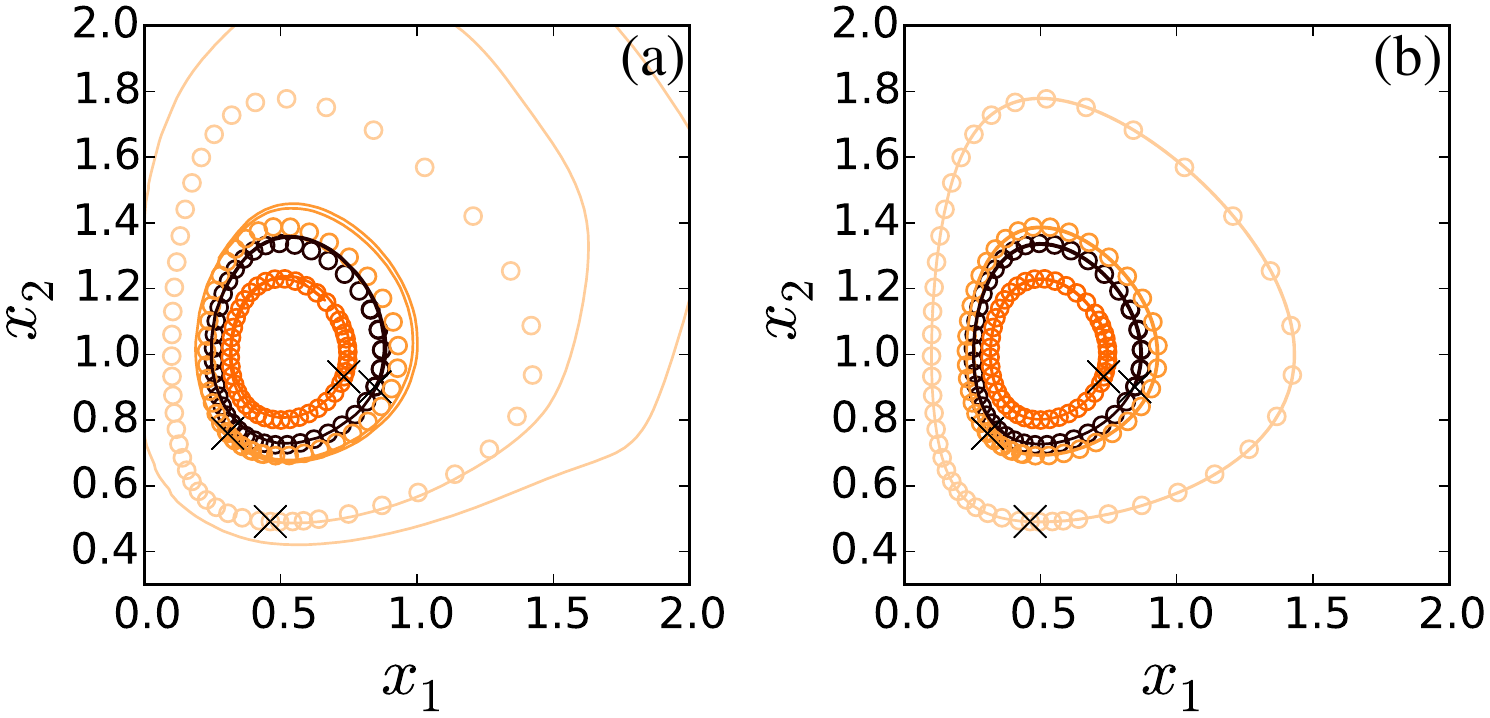}
\end{center}
\caption{Parametric plots of the trajectories $x_1(t)$ and $x_2(t)$
(solid lines) for the (a) inaccurate and (b) accurate
reconstructions of $f_1$ and $f_2$ in 
Figs.~\ref{fi:2d_example_rec_b} and~\ref{fi:2d_example_rec_g},
respectively, for four different initial conditions indicated by the 
crosses. The open circles indicate the trajectories from the 
Lotka-Volterra system of ODEs (Eq.~\ref{eq:mod2}).}
\label{fi:2d_rec_phasespace_gb}
\end{figure}

Using the reconstructions of $f_1$ and $f_2$, we solved for the
trajectories $x_1(t)$ and $x_2(t)$ (for times comparable to $t_{end}$)
and compared them to the trajectories from the original Lotka-Volterra
model (Eq.~\ref{eq:mod2}).  In Fig.~\ref{fi:2d_rec_phasespace_gb} (a)
and (b), we show parametric plots ($x_2(t)$ versus $x_1(t)$) for the
inaccurate (Fig.~\ref{fi:2d_example_rec_b}) and accurate
(Fig.~\ref{fi:2d_example_rec_g}) reconstructions of $f_1$ and $f_2$,
respectively.  We solved both the inaccurate and accurate models with
the same four sets of initial conditions.  For the inaccurate
reconstruction, most of the trajectories from the reconstructed ODE
system do not match the original trajectories.  In fact, some of the
trajectories spiral outward and do not form closed orbits.  In
contrast, for the accurate reconstruction, the reconstructed
trajectories are very close to those of the original model and all possess
closed orbits.

\subsection{Reconstruction of systems of ODEs in 3D}
\label{sec3D}

For the Lorenz ODE model, we need to reconstruct three functions of
three variables: $f_1(x_1,x_2,x_3)$, $f_2(x_1,x_2,x_3)$, and
$f_3(x_1,x_2,x_3)$. Based on the selected parameters $\sigma = 10$,
$\rho = 28$ and $\beta = 8/3$, we chose a $32\times32\times32$
discretization of the domain $-21 \leq x_1 \leq 21$, $-29 \leq x_2
\leq 29$, and $-2 \leq x_3 \leq 50$.  We employed patches of size $10
\times 10$ from each of the $32$ slices (along $x_3$) of size
$32\times 32$ (in the $x_1$-$x_2$ plane) to perform the basis learning.

\begin{figure}[!ht]
\begin{center}
\includegraphics[width=3in]{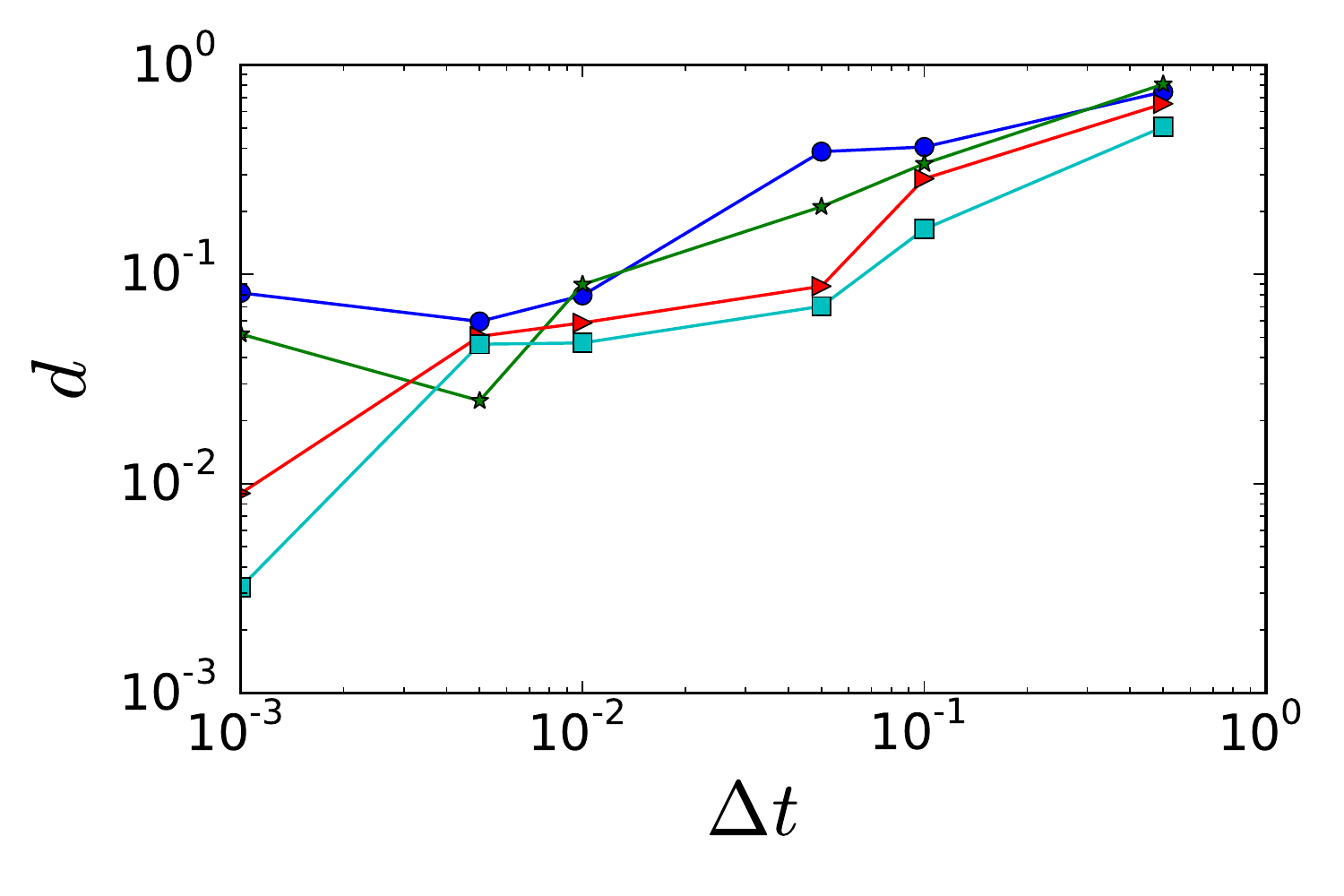}
\end{center}
\caption{Reconstruction error $d$ plotted versus the sampling time
interval $\Delta t$ for the 3D Lorenz model Eq.~(\ref{eq:mod3}) for several 
different numbers of trajectories $N_t = 500$ (circles), $1000$ (stars)
$5000$ (rightward triangles), and $10000$ (squares). Each data 
point is averaged over $20$ independent reconstructions.}
\label{fi:3d_err_vs_dt}
\end{figure}

In Fig.~\ref{fi:3d_err_vs_dt}, we plot the reconstruction error $d$
versus the sampling time interval $\Delta t$ for several $N_t$ from
$500$ to $10^4$ trajectories. As found for the 1D and 2D ODE models,
the reconstruction error decreases with decreasing $\Delta t$ and
increasing $N_t$. $d$ reaches a low-$\Delta t$ plateau that depends on
the value of $N_t$. For $N_t = 10^4$, the low-$\Delta t$ plateau value
for the reconstruction error approaches $d \sim 10^{-3}$. 

\begin{figure*}[!ht]
\includegraphics[width=7in]{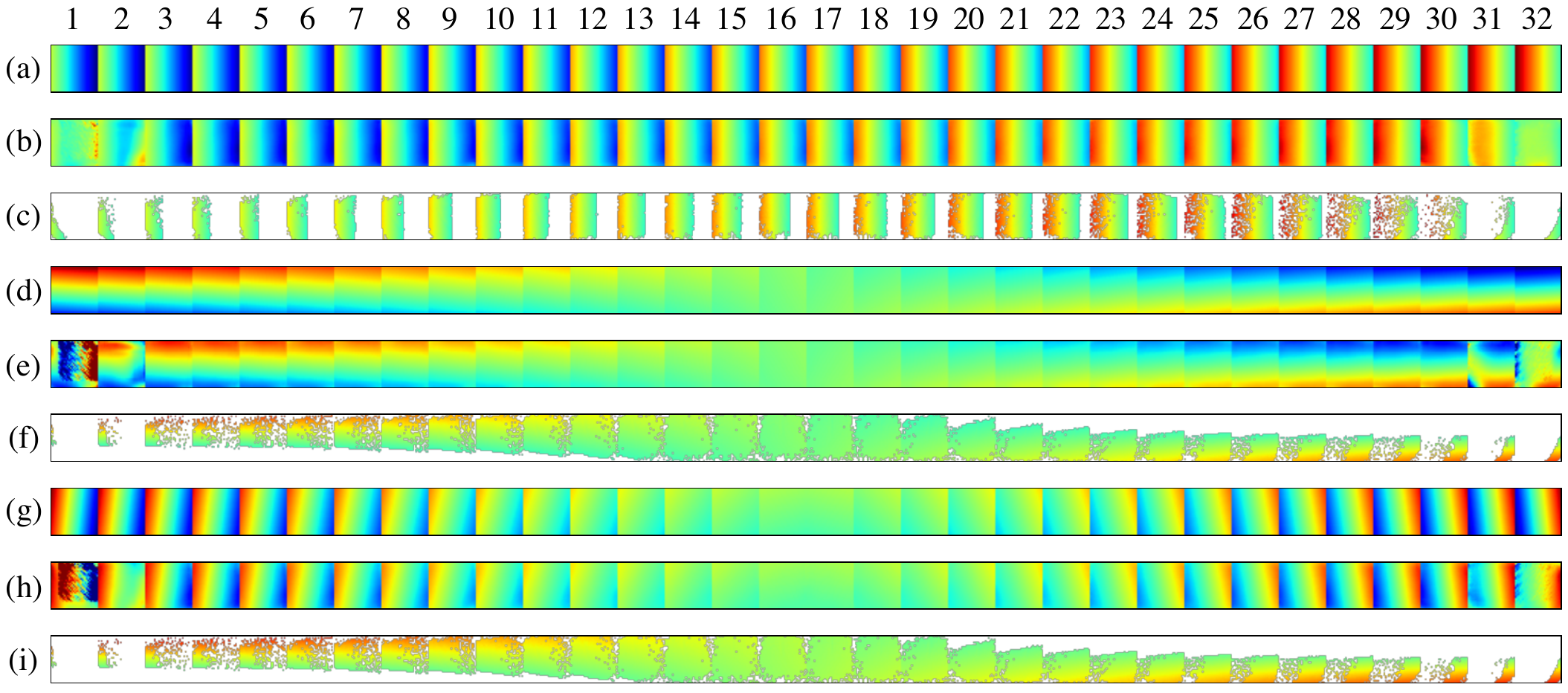}
\caption{Reconstruction of the Lorenz system of ODEs in 3D
(Eq.~\ref{eq:mod3}) using the sampling time interval $\Delta t=
10^{-2}$ and $N_t=10^4$ trajectories.  Panels (a)-(c) indicate
$f_1(i_1)$, (d)-(f) indicate $f_2(i_2)$, and (g)-(i) indicate
$f_3(i_3)$, where the indexes $i_1$, $i_2$, and $i_3$ represent
uniformly spaced $x_1$, $x_2$, and $x_3$ values on the intervals
$-21 \leq x_1 \leq 21$, $-29 \leq x_2 \leq 29$, and $-2 \leq x_3
\leq 50$.  Each of the $32$ panels along $x_3$ represents a $32
\times 32$ discretization of the $x_1$-$x_2$ domain. The first rows 
of each grouping of three ({\it i.e.} panels (a), (d), and (g)) give the
original functions $f_1$, $f_2$, and $f_3$. The second rows 
of each grouping of three ({\it i.e.} panels (b), (e), and (h)) give the 
reconstructed versions of $f_1$, $f_2$, and $f_3$. The third rows of
of grouping of three ({\it i.e.} panels (c), (f), and (i)) show the 
points in the $x_1$, $x_2$, and $x_3$ domain that were used for the 
reconstruction.  The white regions indicate missing data. The color scales range from dark blue to red corresponding to the ranges of $-500 \leq f_1(i_1,i_2,i_3)\leq 500$, $-659 \leq f_2(i_1,i_2,i_3)\leq 659$, and  $-742 \leq f_3(i_1,i_2,i_3)\leq 614$ for the groups of panels (a)-(c), (d)-(f), and (g)-(i), respectively.}
\label{fi:3d_example_rec_g}
\end{figure*}

\begin{figure}[!ht]
\begin{center}
\includegraphics[width=3.5in]{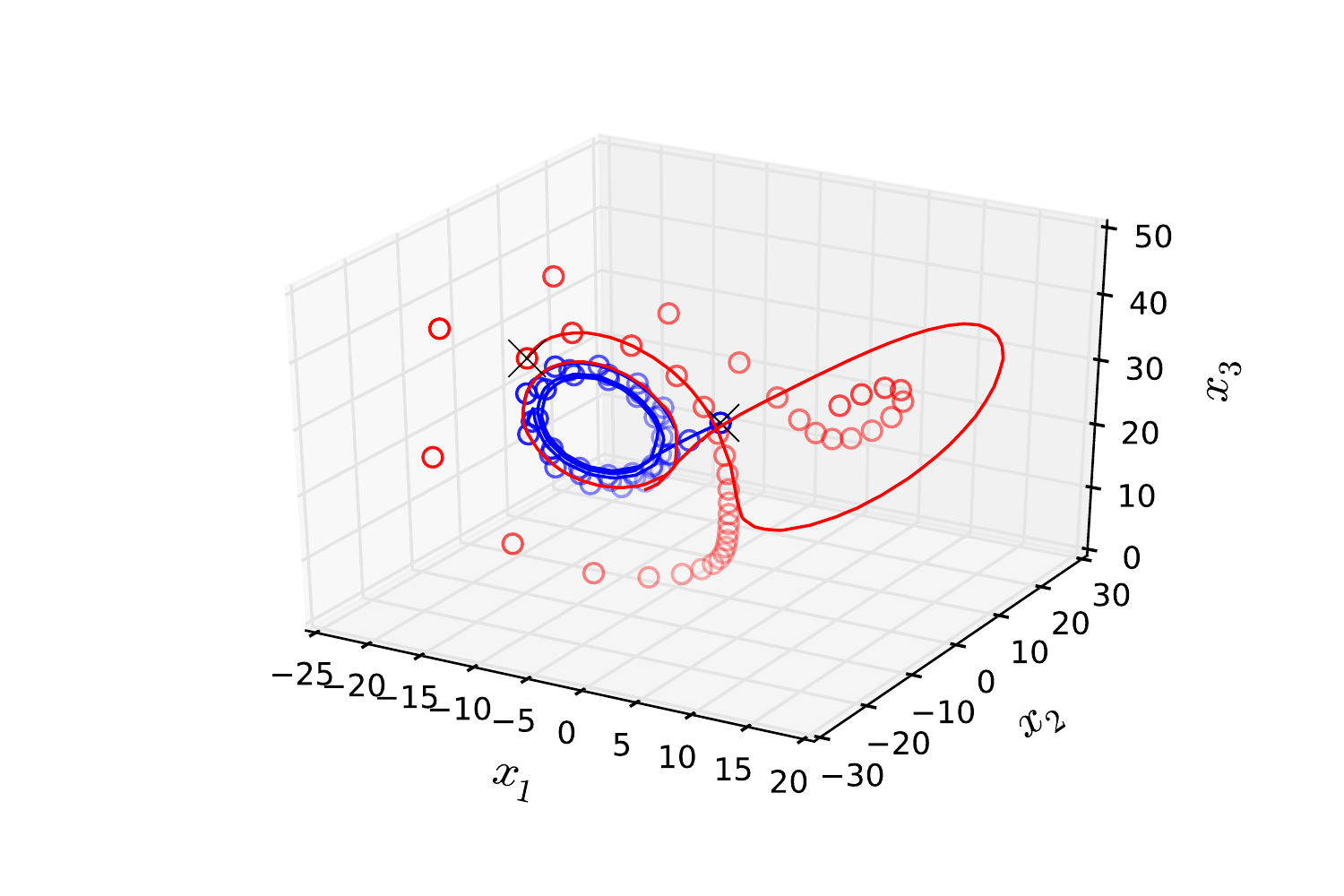}
\end{center}
\caption{We compare the trajectories $x_1(t)$, $x_2(t)$, and $x_3(t)$
from the reconstructed (solid lines) and original (empty circles)
functions $f_1$, $f_2$, and $f_3$ from the 3D Lorenz system of ODEs
(Eq.~\ref{eq:mod3}) with the parameters $\sigma =10$, $\rho=28$, and
$\beta=8/3$ in the chaotic regime.  We plot $x_1$, $x_2$, and $x_3$
parametrically for two initial conditions indicated by the crosses.}
\label{fi:3d_rec_phasespace_g}
\end{figure}

In Fig.~\ref{fi:3d_example_rec_g}, we visualize the reconstructed
functions $f_1$, $f_2$, and $f_3$ for the Lorenz system of
ODEs. Panels (a)-(c) represent $f_1$, (d)-(f) represent $f_2$, and
(g)-(i) represent $f_3$. The 3D domain is broken into $32$ slices
(along $x_3$) of $32\times 32$ grid points in the $x_2$-$x_3$
plane. Panels (a), (d), and (g) give the original functions $f_1$,
$f_2$, and $f_3$ in the Lorenz system of ODEs
(Eq.~\ref{eq:mod3}). Panels (b), (e), and (h) give the reconstructed
versions of $f_1$, $f_2$, and $f_3$, and panels (c), (f), and (i)
provide the data that was used for the reconstructions (with white
regions indicating missing data).  The central regions of the
functions are recovered with high accuracy.  (The edges of the domain
were not well-sampled, and thus the reconstruction was not as
accurate.) These results show that even for chaotic systems in 3D we
are able to achieve accurate ODE reconstruction. In
Fig.~\ref{fi:3d_rec_phasespace_g}, we compare trajectories from the
reconstructed functions to those from the original Lorenz system of
ODEs for times comparable to the inverse of the largest Lyapunov
exponent.  In this case, we find that some of the the trajectories
from the reconstructed model closely match those from the original
model, while others differ from the trajectories of the original
model.  Since chaotic systems are extremely sensitive to initial
conditions, we expect that all trajectories of the reconstructed model
will differ from the trajectories of the original model at long times.
Despite this, the trajectories from the reconstructed model display
chaotic dynamics with similar Lyapunov exponents to those for the
Lorenz system of ODEs, and thus we are able to recover the controlling
dynamics of the original model.

\section{Discussion}
\label{discussion}

We developed a new method for reconstructing sets of nonlinear ODEs
from time series data using machine learning methods involving sparse
function reconstruction and sparse basis learning.  Using only
information from the system trajectories, we first learned a sparse basis,
with no {\it a priori} knowledge of the underlying functions in the
system of ODEs, and then reconstructed the system of ODEs in this
basis.  A key feature of our method is its reliance on sparse
representations of the system of ODEs.  Our results emphasize that
sparse representations provide more accurate reconstructions of
systems of ODEs than least-squares approaches.

We tested our ODE reconstruction method on time series data obtained
from systems of ODEs in 1D, 2D, and 3D.  In 1D, we studied the
Reynolds model for the immune response to infection.  In the parameter
regime we considered, this system possesses only two stable fixed points, and
thus all initial conditions converge to these fixed points in the
long-time limit.  In 2D, we studied the Lotka-Volterra model for
predator-prey dynamics. In the parameter regime we studied, this
system possesses an oscillatory fixed point with closed orbits. In 
3D, we studied the Lorenz model for convective flows.  In the parameter 
regime we considered, the system displays chaotic dynamics on a strange 
attractor. 

For each model, we measured the error in the reconstructed system of
ODEs as a function of parameters of the reconstruction method
including the sampling time interval $\Delta t$, number of
trajectories $N_{t}$, total time $t_{end}$ of the trajectory, and size
of the patches used for basis function learning.  In general, the
error decreases as more data is used for the reconstruction.  We
determined the parameter regimes for which we could achieve highly
accurate reconstruction with errors $d < 10^{-3}$.  We then generated
trajectories from the reconstructed systems of ODEs and compared them
to the trajectories of the original models.  For the 1D model with two
stable fixed points, we were able to achieve extremely accurate
reconstruction and recapitulation of the trajectories of the original
model.  Our reconstruction for the 2D model is also accurate and is
able to achieve closed orbits for most initial conditions.  For some
of the initial conditions, smaller sampling time intervals and longer
trajectories were needed to achieve reconstructed solutions with
closed orbits. In future studies, we will investigate methods to add a
constraint that imposes the constant of the motion on the
reconstruction method, which will allow us to use larger sampling time
intervals and shorter trajectories and still achieve closed orbits.
For the 3D chaotic Lorenz system, we can only match the trajectories
of the reconstructed and original systems for times that are small
compared to the inverse of the largest Lyapunov exponent.  Even though
the trajectories of the reconstructed and original systems will
diverge, we have shown that the reconstructed and original systems of
ODEs possess dynamics with similar Lyapunov exponents.  Now that we
have validated this ODE reconstruction method on known deterministic
systems of ODEs and determined the parameter regimes that yield
accurate reconstructions, we will employ this method in future studies
to identify new systems of ODEs using time series data from
experimental systems for which there is no currently
known system of ODEs.

\acknowledgments 

This work was partially supported by DARPA (Space and Naval Warfare
System Center Pacific) under award number N66001-11-1-4184. These studies
also benefited from the facilities and staff of the Yale University
Faculty of Arts and Sciences High Performance Computing Center and NSF
Grant No. CNS-0821132 that partially funded acquisition of the
computational facilities.

\end{document}